\documentclass[reprint,aps,prb,amsmath,amssymb,twocolumns,showkeys, superscriptaddress,floatfix]{revtex4-2}

\usepackage{silence}
\usepackage{graphicx}% Include figure files
\graphicspath{{./Figures/}}

\usepackage{dcolumn}% Align table columns on decimal point
\usepackage{bm}% bold math
\usepackage[colorlinks = true, allcolors = blue]{hyperref}% add hypertext capabilities
\usepackage{verbatim}
\usepackage{soul}
\usepackage{float}
\usepackage{xcolor}
\usepackage{multirow}
\usepackage{orcidlink}
\usepackage{amsmath}
\usepackage{bbm} %https://tex.stackexchange.com/questions/26637/how-do-you-get-mathbb1-to-work-characteristic-function-of-a-set

\newcommand{\Eq}[1]{Eq.~\eqref{#1}}
\newcommand{\Eqs}[1]{Eqs.~\eqref{#1}}
\newcommand{\Fig}[1]{Fig.~\ref{#1}}

\newcommand{\ket}[1]{\Big| {#1} \Big\rangle}
\newcommand{\braket}[2]{\Big\langle {#1}\Big|{#2}\Big\rangle}

\newcommand{\Real}[1]{\text{Re}\left[ #1 \right]}
\newcommand{\dd}{\mathrm{d}}
\newcommand{\ii}{\mathrm{i}}

\newcommand{\surf}{\parallel}

\newcommand{\dlim}{\displaystyle\lim}
\newcommand{\dint}{\displaystyle\int}

\newcommand{\abs}[1]{\vert #1\vert}
\newcommand{\Det}[1]{\left\vert #1\right\vert}

\begin{document}

\title{Phononic Casimir Effect in Planar Materials}

\author{Pablo Rodriguez-Lopez\,\orcidlink{0000-0003-0625-2682}}
\email{pablo.ropez@urjc.es}
\affiliation{{\'A}rea de Electromagnetismo and Grupo Interdisciplinar de Sistemas Complejos (GISC), Universidad Rey Juan Carlos, 28933, M{\'o}stoles, Madrid, Spain}

\author{Dai-Nam Le\,\orcidlink{0000-0003-0756-8742}}
\email{dainamle@usf.edu}
\affiliation{Department of Physics, University of South Florida, Tampa, Florida 33620, USA}

\author{Lilia M. Woods\,\orcidlink{0000-0002-9872-1847}}
\email{lmwoods@usf.edu}
\affiliation{Department of Physics, University of South Florida, Tampa, Florida 33620, USA}

\date{\today}% It is always \today, today,

\begin{abstract}
The {\it Phononic Casimir effect}   between planar objects is investigated by deriving a formalism from the quantum partition function of the system following multiscattering approach. This  fluctuation-induced coupling is mediated by phonons modeled as an effective elastic medium. We find that excitations with three types of polarizations arise from the resolved boundary conditions, however the coupling is dominated by only one of these degrees of freedom due to exponential suppression effects in the other two. The obtained scaling laws and dependence on materials properties and temperature suggest effective pathways of interaction control. Scenarios of materials combinations are envisioned where the Phononic Casimir effect is of similar order as the standard Casimir interaction mediated by electromagnetic fluctuations.

\end{abstract}

\maketitle

{\it Introduction} Since its theoretical inception as a fluctuation-induced interaction due to the exchange of electromagnetic modes \cite{Casimir1948}, the Casimir force has been established as a universal interaction between objects with fundamental and practical importance \cite{RevModPhys.88.045003,Klimchitskaya2009,Shelden2024}. Unexpected scaling laws and dependence upon fundamental constants in the  Casimir interaction have been found in many novel materials, including topologically nontrivial and anisotropic systems \cite{Rodriguez-Lopez_2023,Le_2022,Broer2023,Muniz2021}. Experimental advances towards new devices, metrology, and measuring the Casimir torque have also become available \cite{Somers2018,Stange2019,Imboden2014}.  Analogous to the electromagnetic Casimir effect, the fluctuations of many observables can also result in similar induced forces. In particular, voltage and charge fluctuations are important for solid state capacitors and biological matter \cite{Drosdoff2016,PARSEGIAN1969,Schoger2022,Naji2010}. Casimir-like interactions have also been found in granular media, systems supporting stochastic processes and critical phenomena \cite{PhysRevE.83.031102,PhysRevLett.96.178001,10.1063/5.0189492,DANTCHEV20231}. The Casimir effect has been proposed for objects in fluids where the fluctuation-induced interaction is mediated by pressure modes. For example, in \cite{Toftul2019,LARRAZA1998151,Reyes2004,Kardar92,DANTCHEV20231} by considering only longitudinal degrees of freedom, the mediated by pressure modes coupling between constant reflectors is considered. Nevertheless, fluctuation induced interactions between objects mediated by acoustic excitations and taking into account their elastic properties and full scattering conditions have not been studied thus far.

The ubiquitous nature of the Casimir force and its relevance to many types of scientific areas motivate further exploration of this widespread phenomenon to other physical realms. It appears that the role of phonons are usually not thought to be important in Casimir phenomena. General considerations imply that the short phonon mean free path and weak phonon-photon coupling render their negligible role in the Casimir force. Nevertheless, recent studies have shown that these general considerations do not reveal the full participation of phonons in fluctuation-induced phenomena. For example, the photon-phonon coupling in piezolectrics can lead to hybrid surface plasmon polaritons responsible for significant modulations of the Casimir force in terms of its sign and magnitude \cite{Le2024}.

In this Letter, we propose the 
{\it Phononic Casimir effect}, a fluctuation-induced interaction mediated by acoustic phonons between objects separated by a solid layer. This is analogous to the standard Casimir interaction from the exchange of electromagnetic modes. The two phenomena have commonalities, however, the Phononic interaction is a consequence of acoustic phonon modes within a condensed matter system, while the electromagnetic coupling does not require any medium in the separating gap between the objects. In principle, only solid materials can host phonons, although there are similarities with properties of fluids especially when the solid materials are considered as elastic media. In particular, the velocity fields characterizing the fluid dynamics can be described by the rate of displacement fields typically used for phonon description in solids. 

\begin{figure}[H]
\centering
\includegraphics[width=0.7\linewidth]{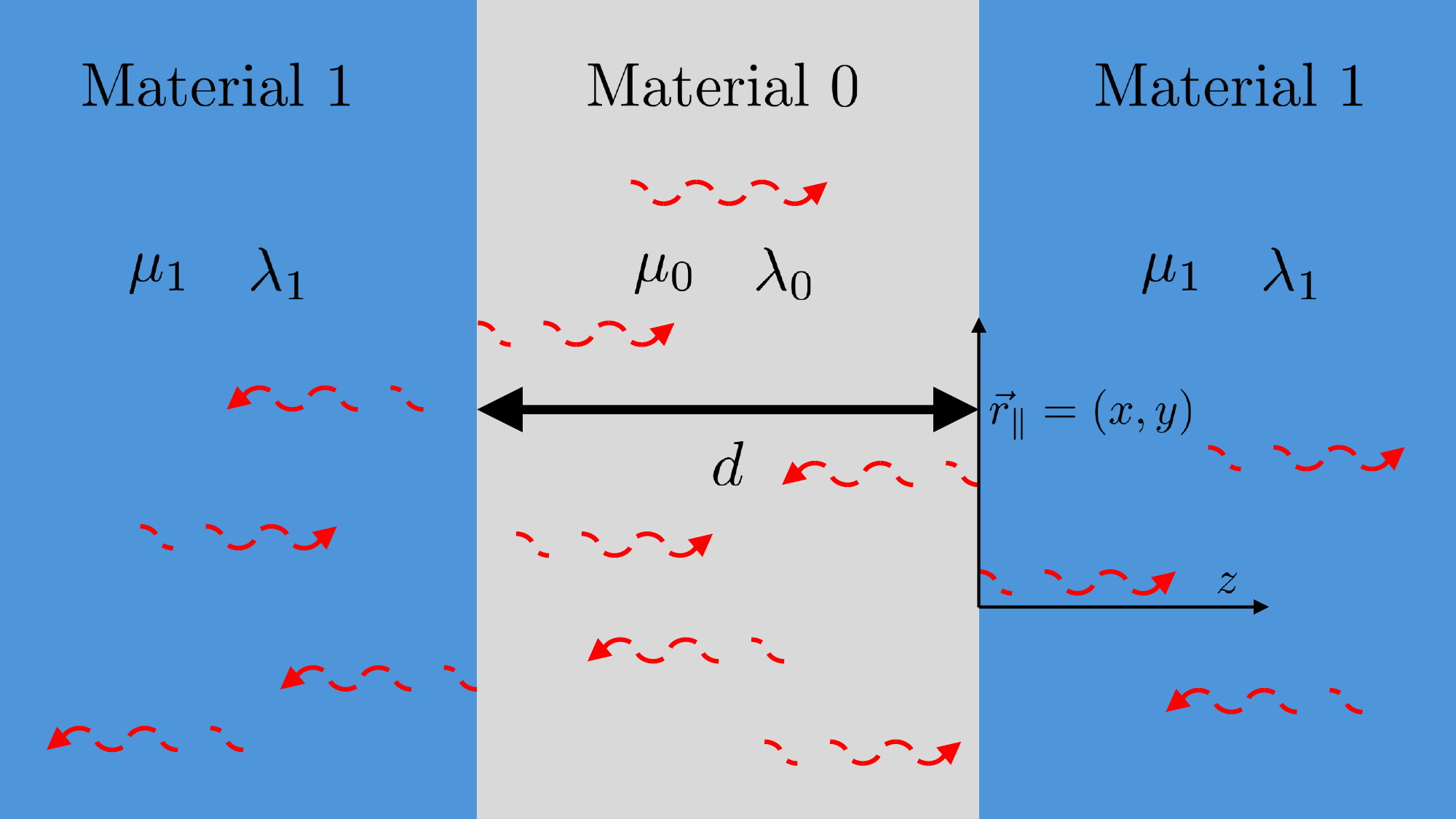}
\caption{Schematics of the considered system where two substrates made of solid Material 1 interact via phonon modes exchange through a gap filled with solid Material 0.}
\label{Fig_PhononicCasimirGeSiGe}
\end{figure}

{\it Theory of the Phononic Casimir effect} The system under consideration (shown in \Fig{Fig_PhononicCasimirGeSiGe})   consists of two planar substrates separated by a gap filled with some material modeled as an elastic continuous medium. We begin with the Cauchy equation of motion, 
\begin{eqnarray}\label{CauchyEq}
\rho\dfrac{D v_{i}}{D t} & = & \partial^{j}\sigma_{ji} + f_{i},
\end{eqnarray}
where the velocity field, $v_{i}(\bm{x},t)$, and the displacement field $u_{i}(\bm{x},t)$ are related as $v_{i}(\bm{x},t) = \partial_{t}u_{i}(\bm{x},t)$). Also, $\sigma_{ji}$ is the stress tensor, $\rho$ is the density of the material, and $f_{i}$ is volumetric density per unit mass of an external force. The  derivative $\dfrac{D v_{i}}{D t}$ tracks the motion of a particle within the elastic medium. For solid materials the lack of advection simplifies the material derivative to ordinary time derivatives $\dfrac{D v_{i}}{D t} = \partial_{t} v_{i} = \partial_{t}^{2}u_{i}$.  As a consequence, the dynamics of the displacement field in solids can be described by suitable  Lagrangian or Hamiltonian formalism \cite{morse1953methods}. The elastic Lagrangian allows us to develop the theoretical framework for the Phononic Casimir effect in analogy to the standard Casimir effect where the electromagnetic action is used as a stepping stone for the definition of the interaction energy. In fact, we can directly apply the well-known tools for the Casimir effect from Quantum Field Theory for electromagnetic fields within a multiscattering formalism \cite{MultiscatteringFormalismEM}. As shown below, from the quantum partition function for acoustic phonon modes under boundary conditions at the interfaces in the considered system, we obtain a generalized Lifshitz-like formula for the phononic Casimir energy. 

The Lagrangian of acoustic phonons within the linear approximation is
\begin{align}\label{Lagrangian_Phonons}
\mathcal{L} = \dfrac{\rho}{2}\dot{u}_{a}\dot{u}^{a} - \frac{1}{2}\sigma_{ab}\varepsilon^{ab} - u_{a}f^{a}
\end{align}
where $u_{a}$ is the displacement from equilibrium position field in the medium, $v_{a} = \dot{u}_{a}$ is its time derivative,  $\varepsilon^{ab} = \frac{1}{2}\big[ \partial^{a}u^{b} + \partial^{b}u^{a} \big]$ is the strain tensor, and $\sigma_{ab}$ is the stress tensor.  The spatial indices are denoted as 
$a,b\in\{x,y,z\}$ and the mass density $\rho$ is assumed constant. In the linear approximation, $\sigma_{ab} = C_{abcd}\varepsilon^{cd}$, where the fourth rank stiffness tensor  $C_{abcd} = \lambda\delta_{ab}\delta_{cd} + \mu\left[ \delta_{ac}\delta_{bd} + \delta_{ad}\delta_{bc}\right]$ for 3D isotropic materials can be expressed in terms of $\lambda$ and $\mu$ being the first and second Lamé parameters.

The quantum partition function of phononic fields in a system with $N$ objects is \cite{MultiscatteringFormalismEM}
\begin{align}
\mathcal{Z} = \int\mathcal{D}\bm{u} \prod_{\alpha=1}^{N}\delta\big[C_{\alpha}[\bm{u}]\big]e^{\frac{\ii}{\hbar}S[\bm{u}]},
\end{align}
where the Dirac delta function $\delta$ enforces the boundary conditions $C_{\alpha}[\bm{u}] = 0$ over the boundary of each $\alpha$-object. Also, $S[\bm{u}] = \int\dd x^{\mu}\mathcal{L}$ is the action of the phononic field associated with the Lagrangian in \Eq{Lagrangian_Phonons}. Following the method in  \cite{MultiscatteringFormalismEM},\cite{lifshitz1956molecular}, the so-obtained   $\mathcal{Z}$  is then utilized for the Casimir free energy between $N$ arbitrary-shaped objects in terms of the bosonic Matsubara frequencies $\omega = \ii\xi_{n} = \ii\frac{2\pi k_{B}T}{\hbar}n$,
\begin{eqnarray}\label{Phononic_Free_Energy}
\mathcal{F} & = & k_{B}T\,\Real{{\sum_{n=0}^{\infty}}'\left(\ln(\mathcal{Z}(\xi_{n})) - \lim_{\abs{\bm{x}_{\alpha\beta}}\to\infty}\mathcal{Z}(\xi_{n})) \right)}\nonumber\\
& = & k_{B}T\,{\sum_{n=0}^{\infty}}'\Real{\ln\Det{\mathbbm{1} - \mathbb{N}(\xi_{n})}},
\end{eqnarray}
where $\ln\Det{\mathbb{A}}$ is the logarithm of the determinant of matrix $\mathbb{A}$. The above expression accounts for removing the partition functions when objects are placed far away $\abs{\bm{x}_{\alpha\beta}}\to\infty$ ensuring a finite result for the interaction energy. The tilde in the sum means that the $n=0$ Matsubara mode is multiplied by $1/2$, and the components of the $\mathbb{N}$ matrix  $\mathbb{N}_{\alpha\beta} = \delta_{\alpha\beta}\mathbb{T}_{\alpha} + [1 - \delta_{\alpha\beta}]\mathbb{U}_{\alpha\beta}$ are related to the scattering matrix $\mathbb{T}_{\alpha}$ and the phonon translation matrix $\mathbb{U}_{\alpha\beta}$ between the $\alpha$ and $\beta$ objects. For  a planar geometry with two interfaces (\Fig{Fig_PhononicCasimirGeSiGe}), \Eq{Phononic_Free_Energy} reduces to 

\begin{align}\label{Lifshitz_formula_finiteT}
\mathcal{F} = k_{B}T{\sum_{n=0}^{\infty}}'\dint_{\mathbb{R}^{2}}\dfrac{\dd^{2}\bm{k}_{\surf}}{(2\pi)^{2}}\Real{\ln\Det{\mathbbm{1} - \mathbb{N}_{12}(\xi_{n})}},
\end{align}
\begin{align}\label{NMatrix}
\mathbb{N}_{12}(\xi_{n}) = \mathbb{R}_{1}(\xi_{n})\mathbb{U}_{12}(\xi_{n})\mathbb{R}_{2}(\xi_{n})\mathbb{U}_{21}(\xi_{n}),
\end{align}
where $\mathbb{R}_{1,2}$ are the  phononic reflection matrices for the two interfaces. 
To obtain the phononic Casimir energy from \Eq{Lifshitz_formula_finiteT}, we  calculate the $\mathbb{R}$ and $\mathbb{U}$ matrices by imposing the boundary conditions for phonon scattering for the planar geometry in Fig. \ref{Fig_PhononicCasimirGeSiGe}, given by 
\begin{eqnarray}\label{HidrostaticBC_ufield}
\sigma_{ij}^{\alpha}n^{j} =  \sigma_{ij}^{\beta}n^{j}, \,\,  \bm{x}\in\Gamma_{\alpha\beta} = \partial\Omega_{\alpha}\cap\partial\Omega_{\beta},
\end{eqnarray}
\begin{eqnarray}\label{NoslipBC_ufield}
\partial_{t}u_{i}^{\alpha} = \partial_{t}u_{i}^{\beta}, \,\,  \bm{x}\in\Gamma_{\alpha\beta} = \partial\Omega_{\alpha}\cap\partial\Omega_{\beta},
\end{eqnarray}
where $n^j$ are the unit vector components of $\mathbf{n}$ perpendicular to the surface of contact $\Gamma_{\alpha\beta}$ between the surface boundaries $\partial\Omega_{\alpha}$ and $\partial\Omega_{\beta}$ of the objects $\alpha$ and $\beta$. The continuity of the deformation at the interface of the two materials due an applied external force in hydrostatic equilibrium is reflected in Eq. \ref{HidrostaticBC_ufield}. There is also a no inter-penetration condition (equivalent to a no-slip boundary conditions in fluids) due to the continuity of the displacement fields across the interface as shown in Eq. \ref{NoslipBC_ufield}.  

To facilitate the calculations, the spectral multipolar decomposition of the displacement field in frequency domain $\bm{u}_{\alpha}(\bm{x},\omega)$ for each object $\alpha$ is sought in terms of multipolar vector wave functions \cite{Hansen1935}\cite{stratton2007electromagnetic} $\bm{L}_{\alpha}(\bm{x},\omega)$, $\bm{M}_{\alpha}(\bm{x},\omega)$ and $\bm{N}_{\alpha}(\bm{x},\omega)$ (more details in Sections I, II in the Supplementary Information \cite{supp}). The $\bm{L}_{\alpha}(\bm{x},\omega)$ function describes a compressional wave (P wave), the $\bm{M}_{\alpha}(\bm{x},\omega)$ function corresponds to shear propagation (SH wave), while the $\bm{N}_{\alpha}(\bm{x},\omega)$ function captures another type of shear modes (SV wave).  By imposing the boundary conditions in Eqs. \ref{HidrostaticBC_ufield},\ref{NoslipBC_ufield}, we obtain the phononic reflection matrix that connects the reflected by the planar boundary multipoles $P_{r}$ with the incident multipoles $Q_{i}$ by $P_{r} = \sum_{Q} R_{P,Q}Q_{i}$ (with $P,Q \in \{L, M, N\}$) as 
\begin{align}
\mathbb{R} = \left(\begin{array}{ccc}
R_{LL} & 0 & R_{LN} \\
%\hline
0 & R_{MM} & 0 \\
R_{NL} & 0 & R_{LN}
\end{array}\right),
\end{align}
where derivation details and the specific form of the coefficients are given in Section III in the Supplementary Information \cite{supp}. Notice that in the electromagnetic Casimir force the reflection matrix for isotropic objects in a planar geometry is a diagonal 2D matrix with separated contributions from transverse electric and transverse magnetic modes \cite{RevModPhys.88.045003,Klimchitskaya2009}. In the Phononic Casimir force  the reflection is a non-diagonal 3D matrix with coefficients involving a mixed contribution from the ${L, N}$-functions. Thus, while the electromagnetic waves polarization is conserved upon scattering at the boundary, this is not the case for the elastic waves. In particular, the $R_{LN}$ coefficients imply that the scattering of longitudinal P waves results into reflected P and SV waves. 

The displacement matrix is also obtained to be of diagonal form
\begin{align}
\mathbb{U} = \left(\begin{array}{ccc}
e^{-d\kappa_{L,z}} & 0 & 0 \\
0 & e^{-d\kappa_{M,z}} & 0 \\
0 & 0 & e^{-d\kappa_{N,z}}
\end{array}\right),
\end{align}
where $d$ is the distance between the substrates and $\kappa_{P,z}=\sqrt{ \left(\frac{\xi}{c_{P}^{2}}\right)^{2} + k_{\surf}^{2} }$ for $\omega = \ii\xi$ for the three modes $P\in\{L,M,N\}$  with their corresponding sound velocities $c_{P}$ ($k_{\surf}$ is the 2D wave vector).

As a first example, we consider the phonon Casimir energy per unit area  between two perfectly reflecting plates in \Fig{Fig_PhononicCasimirGeSiGe}. Thus we impose $\mathbb{R} = \mathbbm{1}_{3\times 3}$, which yields
\begin{eqnarray}\label{phononic_Casimir_energy_vac}
\mathcal{E}_{pr} = - \dfrac{\pi^{2}}{1440}\dfrac{\hbar\left( c_{\ell,0} + 2c_{t,0}\right)}{d^{3}}.
\end{eqnarray}
The above expression has a clear analogy with the Casimir energy between two perfect metals \cite{Casimir1948} $ E_{em}^{pm} = - \dfrac{\hbar c\pi^{2}}{720 d^{3}}$. It appears that the speed of light $c$ is replaced by the sound velocities of the medium in the middle $c_{\ell,0}$ and $c_{t,0}$. The phononic energy takes into account one longitudinal and two transversal phonon polarizations, while the standard Casimir energy is a result of only two transversal polarizations of the electromagnetic field. Since the sound velocity is orders of magnitude smaller than the speed of light ($c = 3\times 10^{8}m/s$ while $c_{\text{sound}} \approx 10^{4}m/s$ ), the phononic energy is much smaller that the electromagnetic one.

Let us now study the interaction between the plates in \Fig{Fig_PhononicCasimirGeSiGe} by taking into account the properties of the materials, as captured in the reflection matrix elements (see Supplementary Information). The specific form of $\mathbb{R}$ allows us to factorize the energy in \Eq{Lifshitz_formula_finiteT} in two separate contributions $\mathcal{E}_{t}=\mathcal{E}_M+\mathcal{E}_{L,N}$, where
\begin{eqnarray}\label{Lifshitz_Formula_fM}
\mathcal{E}_{M} \hspace{-0.1cm}=\hspace{-0.1cm} k_{B}T{\sum_{n=0}^{\infty}}'\hspace{-0.15cm}\dint_{\mathbb{R}^{2}}\hspace{-0.15cm}\dfrac{\dd^{2}\bm{k}_{\surf}}{(2\pi)^{2}}\ln\big[ 1\hspace{-0.1cm} -\hspace{-0.1cm} R_{MM}^{1}R_{MM}^{2}e^{-2d\kappa_{M,z}} \big],
\end{eqnarray}
\begin{eqnarray}\label{Lifshitz_Formula_fLN}
\mathcal{E}_{L,N} \hspace{-0.05cm}=\hspace{-0.05cm} k_{B}T{\sum_{n=1}^{\infty}}\dint_{\mathbb{R}^{2}}\dfrac{\dd^{2}\bm{k}_{\surf}}{(2\pi)^{2}}\Real{\ln\Det{ \mathbbm{1} - \mathbb{N}_{L,N}^{12}(\xi_{n})} },
\end{eqnarray}
\begin{eqnarray}\label{N_Matrix_Lifshitz_Formula_fLN}
\mathbb{N}_{L,N}^{12}(\xi_{n}) \hspace{-0.1cm}=\hspace{-0.1cm} \mathbb{R}_{L,N}^{1}(\xi_{n})\mathbb{U}_{L,N}^{12}(\xi_{n})\mathbb{R}_{L,N}^{2}(\xi_{n})\mathbb{U}_{L,L}^{21}(\xi_{n}).
\end{eqnarray}
In Eq. \ref{Lifshitz_Formula_fM}, $R_{MM}^{1.2}$ are the reflection coefficients for the two interfaces solely for the $M$ phonon polarization for the SH wave. On the other hand, Eq. \ref{Lifshitz_Formula_fLN} is determined by the combined effect of the $LL$ and $LN$ polarizations from the matrix 
$\mathbb{R}_{L,N} = \left(\begin{array}{cc}
R_{LL} & R_{LN}\\
R_{NL} & R_{NN}
\end{array}\right)$ and $\mathbb{U}_{L,N} = \left(\begin{array}{cc}
e^{-d\kappa_{L,z}} & 0\\
0 & e^{-d\kappa_{N,z}}
\end{array}\right)$ giving the combined effects from the P and SV waves.

Assuming that the elastic properties of the materials are independent of frequency and momentum, a change of variables to non-dimensional quantities $\xi = \frac{q}{d}$ and $k_{\surf} = \frac{q_{\surf}}{d}$ in \Eq{Lifshitz_Formula_fLN} results in $\mathbb{R}_{L,N}$ and $\mathbb{U}_{L,N}$ matrices independent of the distance. Thus, one finds that $\mathcal{E}_{L,N}\sim\frac{1}{d^3}$ for all separations. We can obtain an analytical result for this part of the interaction energy by converting 
$k_{B}T{\sum_{n=1}^{\infty}}\rightarrow\frac{\hbar}{2\pi}\int_0^{\infty}d\xi $ in Eq. \ref{Lifshitz_Formula_fLN} and by considering the dominant term in its Laurent series expansion around $\xi=0$ of the $\mathbb{R}_{L,N}$ components found to be proportional to $(\tfrac{k_{\surf}}{\xi})^{2}$. We find that 
\begin{eqnarray}\label{Lifshitz_Formula_Approx2}
\mathcal{E}^{qm}_{L,N} = \dfrac{4 \hbar c_{\ell,0}}{\pi d^{3}}\sqrt{B},
\end{eqnarray}
where the unitless constant $B = \left| \frac{8 \mu_{0} (\mu_{0}-\mu_{1}) }{\lambda_{0} (\mu_{0}+\mu_{1}) + \mu_{0} (\mu_{0}+3 \mu_{1}) } \right| $ is determined by the materials properties, given by the Lamé parameters of the plates $\lambda_{1}$, $\mu_{1}$ and the gap material $\lambda_{0}$, $\mu_{0}$. It turns out that $\mathcal{E}_{L,N}$ is repulsive and it is controlled by the elastic properties. For many  materials the typical range for the constant is $B \sim 0.01-10$ (see Supplementary Information).

 Comparing \Eq{Lifshitz_Formula_Approx2} with  \Eq{phononic_Casimir_energy_vac}, shows that the Phononic Casimir energy due to $L,N$ phonons in the quantum limit is typically bigger than the energy between perfect phonon reflectors. Realizing that the $\mathbb{R}_{L,N}$ matrix elements are proportional to $\xi^{-2}$ for small frequencies shows that these phonon modes are not defined at the $\xi\rightarrow 0$ limit (see Supplementary Information, Section IV \cite{supp}). We find that at $\xi\rightarrow 0$, there is a surface bound state for which no incident wave is possible. This state is always present at the materials interface regardless of the extension of the objects and it is distance-independent.  Therefore, $\xi=0$ is not part of the $L,N$ phononic spectrum contributing to the Casimir effect. This can also be seen from another angle. By considering the $n=0$ term in \Eq{Lifshitz_Formula_fLN}, we find that $\lim_{\xi\to0}\mathcal{E}^{qm}_{L,N}\rightarrow \infty$ is independent of the distance between the plates. Thus, this infinite quantity has no physical meaning and must be removed as part of the energy regularization, as is the case for the electromagnetic Casimir effect \cite{Klimchitskaya2009}. It appears that $L,N$ polarized phonons do not exist at $T=0$ and their phononic Casimir effect must always be considered at finite temperatures. 
 
Based on the above, the Phononic Casimir energy from the $L,N$ modes can be computed using  \Eq{Lifshitz_Formula_fLN} by removing the $n=0$ Matsubara frequency term. Furthermore, we are able to obtain the high temperature limit analytically,
\begin{eqnarray}\label{Regularized_Lifshitz_Formula_Approx_ET2}
\mathcal{E}^T_{L,N} = -\frac{16\hbar c_{\ell,0} B^{2}}{\pi^{2}d^{3}}\left(\dfrac{c_{\ell,0}}{c_{t,0}}\right)^{3}\exp\left(-\frac{4\pi k_{B}Td}{\hbar c_{t,0}}\right).
\end{eqnarray}
Since the sound velocity $c_{t,0}$ is orders of magnitude smaller than $c$, the exponential factor is responsible for a significant decay over a wide distance range, which renders $\mathcal{E}^T_{L,N}$ 
rather unimportant at relevant distances at any temperature.

Further considering the $M$-polarized phonons, the interaction energy at the quantum limit can be obtained by substituting  $k_{B}T{\sum_{n=0}^{\infty}}'\rightarrow\frac{\hbar}{2\pi}\int_0^{\infty}d\xi $ in \Eq{Lifshitz_Formula_fM}, 
\begin{eqnarray}\label{Lifshitz_Formula_fM_T=0}
\mathcal{E}^{qm}_{M} && =  - \dfrac{\hbar c_{t,0}}{16\pi^{2}d^{3}}
\int_{0}^{1}\dd x \text{Li}_{4}\left( \left[\frac{\mu_0 - \mu_1 \gamma(x) }{\mu_0 + \mu_1 \gamma(x) }\right]^{2} \right),
\end{eqnarray}
where $\text{Li}_{4}(u)$ is the  polylogarithm function of order 4 and $\gamma(x) = \sqrt{ \left( \frac{c_{t,0}^{2}}{c_{t,1}^{2}} - 1 \right) x^{2} + 1 }$. We note that all four energies so far, $\mathcal{E}_{pr},  \mathcal{E}_{L,N}^{qm}, \mathcal{E}_{L,N}^T$, and $\mathcal{E}_M^{qm}$, have the same distance dependence and their magnitudes are controlled by the acoustic phonon velocities.  Also, unlike the case of the $L,N$ modes, the $R_{MM}(\xi)$ term is finite for all frequencies meaning that there are no divergencies. 
The high temperature limit is also found, 
\begin{eqnarray}\label{Lifshitz_Formula_fM_highT}
\mathcal{E}^T_{M} = - \frac{k_{B}T}{16\pi d^{2}}\text{Li}_{3}\left( \left[\frac{\mu_{0} - \mu_{1}}{\mu_{0} + \mu_{1}}\right]^{2} \right),
\end{eqnarray}
which is similar to the one obtained in  \cite{Kardar92}, where only longitudinal acoustic modes and perfectly reflecting plates were considered. \Eq{Lifshitz_Formula_fM_highT} also has the same behavior as the high temperature-large distance limit of the electromagnetic Casimir energy between dielectrics $E_{em}^{T} = -\frac{k_{B}T}{16\pi d^{2}}\text{Li}_{3}\left(\left[\frac{\varepsilon_{0} - \varepsilon_{1}}{\varepsilon_{0} + \varepsilon_{1}}\right]^{2}\right)$, which is determined by the zero-frequency dielectric functions of the plates $\epsilon_1$ and the medium between them $\epsilon_0$. Both types of energies have the same scaling law controlled by the dissimilarities in their properties. 

\begin{figure}[H]
\centering
\includegraphics[width=0.9\linewidth]{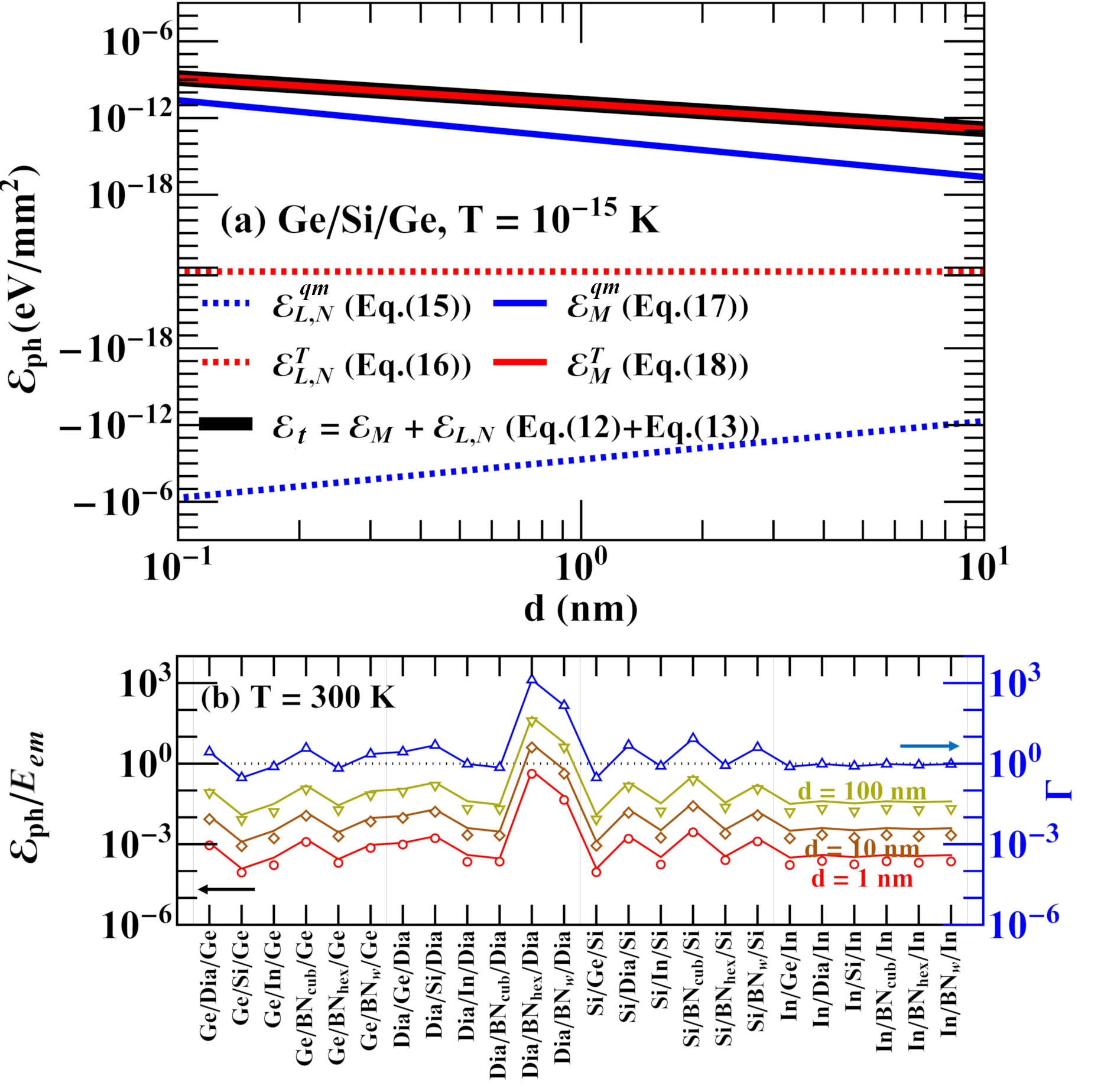}
\caption{(a) Phononic Casimir energy per unit area between Ge plates separated by Si at $T=10^{-15}$ K; (b) Phononic Casimir energy normalized by the electromagnetic Casimir energy for several materials combinations at $T=300$ K. The ratio $\Gamma=\large (\frac{(\mu_0-\mu_1)/(\epsilon_0-\epsilon_2)}{(\mu_0+\mu_1)/(\epsilon_0+\epsilon_2)}\large )^2$ is also shown. The materials parameters are summarized in Table S1 in the Supplementary Information \cite{supp}.}
\label{Fig2}
\end{figure}

{\it Phononic Casimir interactions between materials} 
The importance of the quantum and thermal contributions in the Phononic Casimir energy can be understood in the context of the thermal length $\lambda_T=\frac{\hbar c_{sound}}{k_BT}$ found to be on the order of $0.1$ nm at $T=300$ K. Thus, it is expected that for $d>\lambda_T$ thermal effects dominate the interaction.
As an example of the Phononic Casimir effect with its different contributions, we consider the case of two Ge plates with Si between them at $T=10^{-15}K$ to empathize the importance of the high temperature limit even at such small temperatures.
From the above considerations and results in Fig. \ref{Fig2}(a), we see that there is a wide difference in the  
different types of excitations as part of the energy. At such extremely low temperatures, the contributions from the $L,N$ phonons, quantum and thermal, are several orders of magnitude smaller than the $M$ phonon part. In fact, $\mathcal{E}_{t}$ for separations larger than an Angstrom is practically determined by the thermal $\mathcal{E}_M^T$. This scenario can also be seen in other materials, as shown in the Supplementary Information, Section V \cite{supp}.

The dominance of $\mathcal{E}^T_{M}$ persists at higher temperatures, making the Phononic Casimir energy of classical nature. In Fig. \ref{Fig2}(b), the ratio between $\mathcal{E}^T_{M}$ and the electromagnetic Casimir energy ${E}_{em}$ is shown for several materials combinations at room temperature (detailed calculations are given in the Supplementary Information). We find that  $\mathcal{E}^T_{M}/E_{em}\approx \frac{\pi k_B Td }{\hbar c}\Gamma$ for $d<100$ nm. This is consistent with the results in  Fig. \ref{Fig2}(a), which shows that the Phononic Casimir effect becomes more prominent at higher $T$ and larger $d$. The different contributions for Phononic and electromagnetic Casimir energies  at $T=300$ K are shown for some representative layered materials in the Supplementary Information  (section V) \cite{supp}. The dependence upon $\Gamma$ further indicates that the importance of phonon fluctuations over the electromagnetic fields can be controlled by having materials with very dissimilar Lamé coefficients but dielectric constants of similar values. Experimentally, one can use an AFM cantilever in a direct contact mode or surface force apparatus to measure the Casimir pressure at the Material 1/Material 0 interface for different  $T$ and/or for samples with different $d$ \cite{Alunda2020,OZGUROZER2019,Tonck1991}. Mechanically and thermally controlled transducers \cite{Kellogg2005,MORENO2021} may also be used to measure Young's modules of thin films to indirectly access the Casimir pressure at the interface (see Section VI of the Supplementary Information for details).

In conclusion, we show that fluctuation-induced interaction mediated by phonons gives rise to the Phononic Casimir effect, an analog to the conventional phenomenon from electromagnetic fields. The Phononic Casimir energy is determined by $M$-phonons and it is thermal for practically all separations and temperatures. In practice, both types of contributions must be considered when measuring Casimir energy between planar materials separated by a gap filled with a material. The phononic part can be a substantial part in the experimental observations. By choosing materials with largely dissimilar elastic properties, but very similar dielectric functions,  the phononic energy can become of comparable and even bigger magnitude when compared to the standard electromagnetic energy as demonstrated here. In reality, fluctuation forces have complex origin, and realms beyond electromagnetism need to be considered for accurate description of potential experimental measurements.

{\it Acknowledgments}
P. R.-L. acknowledges support from Ministerio de Ciencia, Innovaci\'on y Universidades (Spain), Agencia Estatal de Investigaci\'on, under project NAUTILUS (PID2022-139524NB-I00). L.M.W. acknowledges financial support from the US Department of Energy under Grant No. DE-FG02-06ER46297.

%\bibliography{bibliography}
%%%%%%%%%%%%%%%%%%%%%%%%%%%%%%%%%%

\providecommand{\noopsort}[1]{}\providecommand{\singleletter}[1]{#1}%
%

%%%%%%%%%%%%%%%%%%%%%%%%%%%%%%%%%%%

\appendix
\begin{widetext}
\section{Supplementary Information: Phononic Casimir Effect in Planar Materials}

\subsection{Basic Equations of Linear Elastic Materials}
The elastic properties of a linear isotropic material are typically quantified by the stress tensor $\sigma$ whose $i,j$-components can be expressed as 
\begin{eqnarray}\label{LinearStressTensor}
\sigma_{ij} = \lambda\varepsilon^{\ell}_{\phantom{\ell}\ell}\delta_{ij} + 2\mu\varepsilon_{ij},
\end{eqnarray}
where  $\lambda$ and $\mu$ are the $1$st and $2$nd Lam\'e's parameters and $\epsilon_{ij}$ are the strain tensor components with $\delta_{ij}$ being the Kronecker delta. The dynamics of a elastic material is governed by the Cauchy's equation of motion given in Eq. (1) of the main text  for the displacement field $\bm{u}(\bm{x},t)$. Taking into account the above relation for a linear isotropic material, the Cauchy's equation of motion takes the form of the Navier-Cauchy equation
\begin{eqnarray}
\rho\ddot{\bm{u}}(\bm{x},t) = \left( \lambda + 2\mu \right)\bm{\nabla}\left(\bm{\nabla}\cdot\bm{u}(\bm{x},t)\right) - \mu\bm{\nabla}\times\bm{\nabla}\times\bm{u}(\bm{x},t) + \rho\bm{f}(\bm{x},t),
\end{eqnarray}
where $\rho$ is the mass density of the material and $\bm{f}$ is the volumetric density per mass unit of an external force. 

By using the Helmholtz decomposition theorem, the displacement field $\bm{u}$ can be decomposed into the sum of an irrotational (curl-free) vector field and a solenoidal (divergence-free) vector field, 
\begin{eqnarray}
\bm{u}(\bm{x},t) = \bm{u}_{\ell}(\bm{x},t) + \bm{u}_{t}(\bm{x},t) = \bm{\nabla}V(\bm{x},t) + \bm{\nabla}\times\bm{A}(\bm{x},t).
\end{eqnarray} 
The irrotational longitudinal displacement field is defined via a scalar potential $V$ satisfying $\bm{\nabla}\times\bm{u}_{\ell} = \bm{\nabla}\times\bm{\nabla}V = \bm{0}$. The solenoidal transversal displacement field can be expressed in terms of a vector potential $\bm{A}$, such that  $\bm{\nabla}\cdot\bm{u}_{t} = \bm{\nabla}\cdot\bm{\nabla}\times\bm{A} = 0$. Similar decomposition can be applied to the volumetric forces $\bm{f} = \bm{f}_{\ell} + \bm{f}_{t}$ with properly chosen scalar and vector potentials. 

The dynamical Navier-Cauchy equation for the isotropic linear material can then be separated into longitudinal and transversal parts:
\begin{eqnarray}
\ddot{\bm{u}}_{\ell}(\bm{x},t) & = & c_{\ell}^{2}\Delta\bm{u}_{\ell}(\bm{x},t) + \bm{f}_{\ell}(\bm{x},t),
\end{eqnarray}
\begin{eqnarray}
\ddot{\bm{u}}_{t}(\bm{x},t) & = & c_{t}^{2}\Delta\bm{u}_{t}(\bm{x},t) + \bm{f}_{t}(\bm{x},t),
\end{eqnarray}
where the longitudinal and transverse sound velocities are related to the Lam\'e coefficients, $c_{\ell} = \sqrt{ \frac{\lambda + 2\mu}{\rho} }$ and $c_{t} = \sqrt{ \frac{\mu}{\rho} }$  \cite{auld1973acoustic1,morse1953methods}.

The displacement field in the elastic material is analogous to the electromagnetic fields governed by Maxwell's equations. In electromagnetism, the modes contributing to the standard Casimir effect are of transverse polarization only with separable transverse electric (TE) and transverse magnetic (TM) excitations in planar isotropic media. For elastic fields, however, there are not only two types of transverse excitations, but there are longitudinal modes which also contribute to the Phononic Casimir effect. This implies that the fluctuation-induced phenomenon mediated by phonons is a generalization of the electromagnetic theory used for standard Casimir interactions. In what follows, we show explicitly the emergence of two transverse and one longitudinal phonon modes using spectral multipolar functions satisfying elastic boundary conditions in layered isotropic materials.

\subsection{Spectral multipolar decomposition}
To find the solutions of the Navier-Stokes equation, the displacement field $\bm{u}(\bm{x},t)$ is represented in terms of a  multipolar expansion using vector wave functions
$\bm{L}(\bm{x},t)$ (corresponding to longitudinal polarization) and $\bm{M}(\bm{x},t), \bm{N}(\bm{x},t)$ (corresponding to two transverse polarizations),
\begin{eqnarray}
\bm{u}(\bm{x},t) = \bm{u}_{\ell}(\bm{x},t) + \bm{u}_{t}(\bm{x},t) = u_{L}\bm{L}(\bm{x},t) + \Big[ u_{M}\bm{M}(\bm{x},t) + u_{N}\bm{N}(\bm{x},t) \Big].
\end{eqnarray}
Starting from the scalar plane wave equation
\begin{eqnarray}
\ddot{\phi}_{a}(\bm{x},t) = c_{a}^{2}\Delta\phi_{a}(\bm{x},t),
\end{eqnarray}
whose solution in free space is given in Cartesian coordinates as $\phi_{a} = e^{\ii\left( \bm{k}\cdot\bm{x} - \omega t \right)}$ with $\omega^{2} = c_{a}^{2}k_{j}k^{j} = c_{a}^{2}\bm{k}^{2}$, the multipolar functions are given as 
\begin{eqnarray}
\bm{L} & = & \frac{1}{k^{2}}\bm{\nabla}\phi_{\ell}, \nonumber\\
\bm{M} & = & \dfrac{1}{k_{\surf}}\bm{\nabla}\times\left[\phi_{t}\hat{\bm{z}}\right], \nonumber\\
\bm{N} & = & \dfrac{1}{k}\bm{\nabla}\times\bm{M},
\end{eqnarray}
where $\bm{k} = (k_{x},k_{y},k_{z})$, $\bm{k}_{\surf} = (k_{x},k_{y})$, $k_{\surf} = \sqrt{ k_{x}^{2} + k_{y}^{2} }$, and $k = \sqrt{ k_{\surf}^{2} + k_{z}^{2} }$, and $\hat{\bm{z}}$ is the constant vector field along z-axis. Consequently, we have 
\begin{eqnarray}
\begin{array}{ccc}
\partial_{t}^{2}\bm{L} = c_{\ell}^{2}\Delta\bm{L} &
\partial_{t}^{2}\bm{M} = c_{t}^{2}\Delta\bm{M} &
\partial_{t}^{2}\bm{N} = c_{t}^{2}\Delta\bm{N}\vspace{0.2cm}\\
\bm{M} = \dfrac{c_{t}}{\omega}\bm{\nabla}\times\bm{N} &
\omega_{\ell}^{2} = c_{\ell}^{2}k^{2} &
\omega_{t}^{2} = c_{t}^{2}k^{2}
\end{array}
\end{eqnarray}
with $u_{P} =\bm{P}^{\dagger}\cdot\bm{u} =\int\dd t\int_{\Omega}\dd\bm{x}\bm{P}^{\dagger}(\bm{x},t)\cdot\bm{u}(\bm{x},t)$, being $P = \{L,M, N \}$.

%\begin{eqnarray}
%\bm{L}\cdot\bm{L}^{\dagger} = \bm{M}\cdot\bm{M}^{\dagger} = \bm{N}\cdot\bm{N}^{\dagger} = 1\\
%\bm{L}\cdot\bm{M}^{\dagger} = \bm{L}\cdot\bm{N}^{\dagger} = \bm{M}\cdot\bm{N}^{\dagger} = 0
%\end{eqnarray}

\subsection{Reflection and Transmission matrices}
In this section, we derive the reflection and transmission coefficients for two thick isotropic  plates in contact and in mechanical equilibrium. The two plates are made of isotropic elastic materials. 
 We start by constructing the phononic fields in the two different media denoted as $\alpha$ and $\beta$ (see Fig. \ref{Fig_Fresnel_Problem}): an incident wave $\bm{u}_{i}$ in $\alpha$ ($z<0$) strikes the planar boundary with materials $\beta$, which produces a reflected wave $\bm{u}_{r}$ in material $\alpha$ and a transmitted wave $\bm{u}_{t}$ into material $\beta$ ($z>0$). These are further decomposed using the above described multipolar expansion in terms of the $\bm{L}(\bm{x},t), \bm{M}(\bm{x},t), \bm{N}(\bm{x},t)$ vector wave functions,

\begin{eqnarray}\label{Scattering_Ansatz}
\bm{u}_{i} & = & A_{L,i}\bm{L}_{\alpha}(\bm{k}_{i,L},\bm{r}) + \Big[ A_{M,i}\bm{M}_{\alpha}(\bm{k}_{i,M},\bm{r}) + A_{N,i}\bm{N}_{\alpha}(\bm{k}_{i,N},\bm{r}) \Big],\nonumber\\
\bm{u}_{r} & = & A_{L,r}\bm{L}_{\alpha}(\bm{k}_{r,L},\bm{r}) + \Big[ A_{M,r}\bm{M}_{\alpha}(\bm{k}_{r,M},\bm{r}) + A_{N,r}\bm{N}_{\alpha}(\bm{k}_{r,N},\bm{r}) \Big],\nonumber\\
\bm{u}_{t} & = & A_{L,t}\bm{L}_{\beta}(\bm{k}_{t,L},\bm{r}) + \Big[ A_{M,t}\bm{M}_{\beta}(\bm{k}_{t,M},\bm{r}) + A_{N,t}\bm{N}_{\beta}(\bm{k}_{t,N},\bm{r}) \Big].
\end{eqnarray}

To generate the vector multipolar functions, we  use the scalar solution to Eq. (7) 

\begin{eqnarray}
\phi_{\alpha,P}(\bm{k}_{i,P},\bm{x}) = e^{\ii\left( \bm{k}_{\surf}\cdot\bm{r}_{\surf} + k_{P}z - \omega t \right)},\hspace{0.3cm}
\phi_{\alpha,P}(\bm{k}_{r,P},\bm{x}) = e^{\ii\left( \bm{k}_{\surf}\cdot\bm{r}_{\surf} - k_{P}z - \omega t \right)},\hspace{0.3cm}
\phi_{\beta,P}(\bm{k}_{t,P},\bm{x}) = e^{\ii\left( \bm{k}_{\surf}\cdot\bm{r}_{\surf} + q_{P}z - \omega t \right)},
\end{eqnarray}
where $\bm{k}_{\surf}\cdot\bm{r}_{\surf} =  k_{x}x + k_{y}y$, $P\in\{L,M,N\}$ denotes the specific vector multipole and $\bm{k}_{i,P}, \bm{k}_{r,P},$ and $\bm{k}_{t,P}$ are the wave vectors for their incident, reflected, and transmitted wave vectors, respectively. Note that we have used that the frequency (therefore, the energy) is conserved $\omega = \omega_{i} = \omega_{r} = \omega_{t}$ and that the problem is invariant in the $(x,y)$ plane, therefore $k_{x} = k_{x,i} = k_{x,r} = k_{x,t}$ and $k_{y} = k_{y,i} = k_{y,r} = k_{y,t}$. The $z$-components of the wave vectors in the exponential factors contain the phonon acoustic velocities, 
\begin{eqnarray}
k_{P} = \sqrt{ \left(\frac{\omega}{c_{P,\alpha}^{2}}\right)^{2} - k_{\surf}^{2} },\hspace{2cm}
q_{P} = \sqrt{ \left(\frac{\omega}{c_{P,\beta}^{2}}\right)^{2} - k_{\surf}^{2} },
\end{eqnarray}
taking into account one longitudinal ($P=\{L\}$) and two transverse ($P=\{M,N\}$) polarizations. For isotropic materials, the two transversal waves ($\bm{M}$ and $\bm{N}$ vector multipoles) have the same velocity $c_{M} = c_{N}$. It is also important to note that despite that the longitudinal and transversal waves have the same frequency $\omega$ and the same momentum components $k_{x}$ and $k_{y}$, they cannot have the same direction ($k_{L}\neq k_{N})$ because such waves travel at different velocities ($c_{L}\neq c_{N}$). The schematics of the incident, reflected, and transmitted waves in system composed of the thick planar substrates in contact is shown in Fig. \ref{Fig_Fresnel_Problem}.

\begin{figure}[H]
\centering
\includegraphics[width=0.45\linewidth]{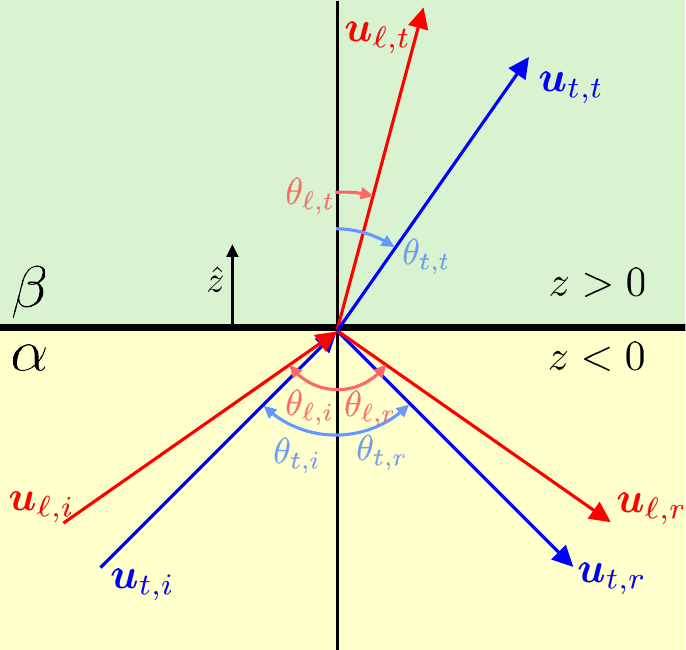}
\caption{Schematics of the incident, reflected, and transmitted elastic waves in two isotropic media with a shared planar boundary. Blue arrows represent transversal waves with $\bm{M}$ corresponding to an SH wave and $\bm{N}$ corresponding to an SV wave. Red arrows represent longitudinal waves labeled as P waves, given by the $\bm{L}$ functions. The angles that the incident, reflected, and transmitted elastic waves with the transverse and longitudinal polarizations make with the vertical are also denoted. 
The reflected and transmitted planar waves have the same frequency $\omega$ and parallel momentum $\bm{k}_{\surf} = (k_{x},k_{y})$, but they have  different sound velocities, thus $k_{z,\ell} \neq k_{z,t}$).}
\label{Fig_Fresnel_Problem}
\end{figure}

To find the relation between the incident, reflected and transmitted waves, we utilize the no-slip boundary condition (Eq. (8) in the main text). This is a fundamental assumption in fluid dynamics with relevance to phonons. It assumes that the elastic waves do not break the connection between the adjacent surfaces at their shared boundary, and it is described as 
\begin{eqnarray}\label{No_Slip_BC_Appendix}
\lim_{z\to 0^{+}}\partial_{t}\bm{u}^{\alpha}(z) = \lim_{z\to 0^{-}}\partial_{t}\bm{u}^{\beta}(z).
\end{eqnarray}
To continue further, we project \Eq{No_Slip_BC_Appendix} to the polarization basis by multiplying this boundary condition by the vector functions basis
$\bm{L}^{\dagger}(\bm{k}_{i,L},\bm{r})$, $\bm{M}^{\dagger}(\bm{k}_{i,L},\bm{r})$ and $\bm{N}^{\dagger}(\bm{k}_{i,L},\bm{r})$  with $\bm{k}_{i,L} = (k_{x}, k_{y}, k_{L})$. This leads to three auxiliary equations summarized below, where each line represent the case of $P=\{L,M,N\}$ respectively, 
\begin{eqnarray}\label{NoSlipBC_expanded_alpha}
\begin{array}{rcl}
\dlim_{z\to 0^{+}}\dfrac{\ii\omega}{c_{L,\alpha}}\bm{P}^{\dagger}(\bm{k}_{i,L},\bm{r})\cdot\bm{u}^{\alpha}(\bm{r})
& = & \left\lbrace
\begin{array}{l}
\frac{\omega^{2} A_{L,i}}{c_{L,\alpha}}+c_{L,\alpha} \left(k_{\surf}^{2}-k_{L}^{2}\right) A_{L,r}-\ii c_{N,\alpha} k_{\surf} \left((k_{L}-k_{N})
   A_{N,i}+(k_{L}+k_{N}) A_{N,r}\right)\\
A_{M,i}+A_{M,r}\\
c_{N,\alpha} \left(\left(k_{\surf}^{2} + k_{L} k_{N}\right) A_{N,i}+\left(k_{\surf}^{2}-k_{L} k_{N}\right) A_{N,r}\right) - 2\ii c_{L,\alpha} k_{L} k_{\surf} A_{L,r}
\end{array}\right.\\
\\
\dlim_{z\to 0^{-}}\dfrac{\ii\omega}{c_{L,\alpha}}\bm{P}^{\dagger}(\bm{k}_{i,L},\bm{r})\cdot\bm{u}^{\beta}(\bm{r}) & = & \left\lbrace
\begin{array}{l}
c_{L,\beta} \left(k_{L} q_{L}+k_{\surf}^{2}\right) A_{L,t} + \ii c_{N,\beta} k_{\surf} (q_{N}-k_{L}) A_{N,t}\\
A_{M,t}\\
c_{N,\beta} \left(k_{L} q_{N}+k_{\surf}^{2}\right) A_{N,t} + \ii c_{L,\beta} k_{\surf} (q_{L}-k_{L}) A_{L,t}
\end{array}\right.
\end{array}
\end{eqnarray}

An additional boundary condition, which gives further relations between the incident, reflected, and transmitted waves, is the continuity of deformation from the local balance of momenta equation (Eq. (7) in the main text with $\hat{\bm{n}} = \hat{\bm{z}}$),
\begin{eqnarray}\label{HidrostaticEqBC_expanded}
\lim_{z\to 0^{+}}\sigma^{\alpha}_{i3}(z) = \lim_{z\to 0^{-}}\sigma^{\beta}_{i3}.
\end{eqnarray}
Projecting the stress tensor components onto the $\bm{L}^{\dagger}(\bm{k}_{i,L},\bm{r})$, $\bm{M}^{\dagger}(\bm{k}_{i,L},\bm{r})$ and $\bm{N}^{\dagger}(\bm{k}_{i,L},\bm{r})$ functions, additional three equations are obtained 

\begin{eqnarray}\label{HidrostaticBC_expanded_alpha}
\begin{array}{rcl}
\dlim_{z\to 0^{-}}\dfrac{\omega^{2}}{c_{L,\alpha}}P^{\dagger,j}(\bm{k}_{i,L},\bm{r})\sigma^{\alpha}_{j3}(\bm{r}) & = & \left\lbrace
\begin{array}{l}
\frac{\ii k_{L} \omega^{2} \lambda_{\alpha} \left(A_{L,i}+A_{L,r}\right)}{c_{L,\alpha}} + 2 \ii k_{L} \mu_{\alpha} \left(\frac{\omega^{2} A_{L,i}}{c_{L,\alpha}}+c_{L,\alpha} \left(k_{L}^{2}-k_{\surf}^{2}\right) A_{L,r}\right)\\
\hspace{1cm}+ c_{N,\alpha} k_{\surf} \mu_{\alpha}\left(\left( k_{\surf}^{2} - k_{N}^{2} + 2 k_{L} k_{N}\right) A_{N,i}+\left( k_{\surf}^{2} - k_{N}^{2} - 2 k_{L} k_{N} \right)A_{N,r}\right)\\
k_{M} \mu_{\alpha} \left(A_{M,i}-A_{M,r}\right)\\
-\frac{k_{\surf} \omega^{2} \lambda_{\alpha} \left(A_{L,i}+A_{L,r}\right)}{c_{L,\alpha}}-4 c_{L,\alpha} k_{L}^{2} k_{\surf} \mu_{\alpha}A_{L,r}-\ii c_{N,\alpha} \mu_{\alpha} \Big(\left(k_{L} \left(k_{\surf}^{2}-k_{N}^{2}\right)-2 k_{N} k_{\surf}^{2}\right)A_{N,i}\\
\hspace{1cm}+\left(k_{L} \left(k_{\surf}^{2}-k_{N}^{2}\right)+2 k_{N} k_{\surf}^{2}\right) A_{N,r}\Big)
\end{array}\right.\\
\\
\dlim_{z\to 0^{+}}\dfrac{\omega^{2}}{c_{L,\alpha}}P^{\dagger,j}(\bm{k}_{i,L},\bm{r})\sigma^{\beta}_{j3}(z) & = & \left\lbrace
\begin{array}{l}
\frac{\ii k_{L} \omega^{2} \lambda_{\beta} A_{L,t}}{c_{L,\beta}} + \mu_{\beta} \left(c_{N,\beta} k_{\surf} \left( k_{\surf}^{2} - q_{N}^{2} + 2 k_{L} q_{N}\right) A_{N,t}+2 \ii c_{L,\beta} q_{L} \left(k_{L}q_{L}+k_{\surf}^{2}\right) A_{L,t}\right)\\
q_{M} \mu_{\beta} A_{M,t}\\
-\frac{k_{\surf} \omega^{2} \lambda_{\beta} A_{L,t}}{c_{L,\beta}}+\mu_{\beta} \left(2 c_{L,\beta} k_{\surf} q_{L} (k_{L}-q_{L}) A_{L,t} - \ii c_{N,\beta} \left(k_{L} \left(k_{\surf}^{2}-q_{N}^{2}\right)-2 k_{\surf}^{2} q_{N}\right) A_{N,t}\right)
\end{array}\right.
\end{array}
\end{eqnarray}

Solving \Eqs{NoSlipBC_expanded_alpha} and \Eqs{HidrostaticBC_expanded_alpha} by expressing the unknown reflected and transmitted amplitudes in terms of the incident fields, we derive the phononic reflection and transmission matrices as,
\begin{eqnarray}
\bm{A}_{r} = \left(\begin{array}{c}
A_{L,r}\\
A_{M,r}\\
A_{N,r}
\end{array}
\right) & = & R\bm{A}_{0} = \left(\begin{array}{ccc}
R_{LL} & 0 & R_{LN}\\
0& R_{MM} & 0\\
R_{NL} & 0 & R_{LL}
\end{array}
\right)\left(\begin{array}{c}
A_{L,i}\\
A_{M,i}\\
A_{N,i}
\end{array}
\right)
\end{eqnarray}
\begin{eqnarray}
\bm{A}_{t} = \left(\begin{array}{c}
A_{L,t}\\
A_{M,t}\\
A_{N,t}
\end{array}
\right) & = & T\bm{A}_{0} = \left(\begin{array}{ccc}
T_{LL} & 0 & T_{LN}\\
0& T_{MM} & 0\\
T_{NL} & 0 & T_{LL}
\end{array}
\right)\left(\begin{array}{c}
A_{L,i}\\
A_{M,i}\\
A_{N,i}
\end{array}
\right)
\end{eqnarray}

The structure of these $3\times 3$ matrices shows that the boundary conditions preserve the $M$ transverse polarization, however, there is mixing between the $N$ transverse polarization and the $L$ longitudinal phonons as evident in the emergence of $R_{LN},R_{NL}$ components. 

The explicit expressions for the different coefficients are given below starting with the reflection for the $M$ phonons,

\begin{eqnarray}
R_{MM} = \frac{\mu_{\alpha}k_{M} - \mu_{\beta}q_{M}}{\mu_{\alpha}k_{M} + \mu_{\beta}q_{M}},
\end{eqnarray}

\begin{eqnarray}
T_{MM} = \frac{2\mu_{\alpha}k_{M}}{\mu_{\alpha}k_{M} + \mu_{\beta}q_{M}}.
\end{eqnarray}

For the reflected  $L$-polarized waves, we find 

\begin{eqnarray}
\Delta R_{LL} & = & -2 \Bigg[\mu_{\alpha} \mu_{\beta} \Bigg\lbrace\omega^{4} \left(\frac{k_{N} q_{L}}{c_{L,\alpha}^{2} c_{N,\beta}^{2}}-\frac{k_{L} q_{N}}{c_{L,\beta}^{2} c_{N,\alpha}^{2}}\right)\nonumber\\
& &+k_{\surf}^{2}\left[k_{L}^{2} \left(k_{\surf}^{2}+2 q_{L} q_{N}\right)+k_{L} k_{N} \left(k_{\surf}^{2}+3 q_{L} q_{N}\right) -(k_{L}+k_{N}) \left(k_{L} q_{N}^{2}+k_{N} q_{L}^{2}\right)+ q_{L}q_{N}\left(k_{N}^{2} - k_{\surf}^{2}\right) + k_{\surf}^{2}q_{L}^{2}\right]\Bigg\rbrace\nonumber\\
& &-k_{L} \mu_{\alpha}^{2} \left(k_{\surf}^{2} (k_{L}+2 k_{N})-k_{L} k_{N}^{2}\right) \left(k_{\surf}^{2}+q_{L} q_{N}\right)+q_{L} \mu_{\beta}^{2}\left(k_{L} k_{N}-k_{\surf}^{2}\right) \left(k_{\surf}^{2} (q_{L}-2 q_{N})-q_{L} q_{N}^{2}\right)\Bigg]\nonumber\\
& &+\frac{\omega^{2} \lambda_{\beta} \left(\mu_{\alpha} \left(\frac{k_{L}q_{N} \omega^{2}}{c_{N,\alpha}^{2}}+k_{\surf}^{2} \left(2 k_{L} k_{N}+k_{N}^{2}-k_{\surf}^{2}\right)\right) + \mu_{\beta}\left(q_{N}^{2}-k_{\surf}^{2}\right) \left(k_{L}k_{N}-k_{\surf}^{2}\right)\right)}{c_{L,\beta}^{2}}\nonumber\\
& &-\frac{\omega^{2} \lambda_{\alpha} \left(\mu_{\beta} \left(\frac{k_{N} q_{L} \omega^{2}}{c_{N,\beta}^{2}}+k_{\surf}^{2} \left( 2q_{L} q_{N} - q_{N}^{2} + k_{\surf}^{2}\right)\right) + \mu_{\alpha}\left(k_{N}^{2}-k_{\surf}^{2}\right) \left(q_{L} q_{N} + k_{\surf}^{2}\right)\right)}{c_{L,\alpha}^{2}},
\end{eqnarray}
where 
\begin{eqnarray}\label{Eq_Delta}
\Delta & = & 2 \Bigg[\mu_{\alpha} \mu_{\beta} \left\lbrace\omega^{4} \left(\frac{k_{L} q_{N}}{c_{L,\beta}^{2} c_{N,\alpha}^{2}}+\frac{k_{N} q_{L}}{c_{L,\alpha}^{2} c_{N,\beta}^{2}}\right) + k_{\surf}^{2} \left[k_{L}^{2} \left(k_{\surf}^{2}-q_{N}^{2}+2 q_{L} q_{N}\right)+k_{L}k_{N} \left(2 q_{L}^{2}+q_{N}^{2}-k_{\surf}^{2}-4 q_{L} q_{N}\right) \right]\right\rbrace\nonumber\\
& & +k_{L} \mu_{\alpha}^{2} \left(k_{L} \left(k_{N}^{2}-k_{\surf}^{2}\right)+2 k_{N} k_{\surf}^{2}\right) \left(k_{\surf}^{2}+q_{L}q_{N}\right) + q_{L} \mu_{\beta}^{2} \left( q_{L} \left(q_{N}^{2} - k_{\surf}^{2}\right) + 2 k_{\surf}^{2} q_{N}\right)\left(k_{\surf}^{2} + k_{L} k_{N}\right)\Bigg]\nonumber\\
& & +\frac{\omega^{2} \lambda_{\beta} \left(\mu_{\alpha} \left(\frac{k_{L} q_{N} \omega^{2}}{c_{N,\alpha}^{2}}+k_{\surf}^{2} \left(k_{\surf}^{2}+2 k_{L}k_{N}-k_{N}^{2}\right)\right) + \mu_{\beta}\left(q_{N}^{2} - k_{\surf}^{2}\right)  \left(k_{\surf}^{2} + k_{L} k_{N}\right)\right)}{c_{L,\beta}^{2}}\nonumber\\
& &+\frac{\omega^{2} \lambda_{\alpha} \left(\mu_{\beta} \left(\frac{k_{N} q_{L} \omega^{2}}{c_{N,\beta}^{2}}+k_{\surf}^{2} \left(k_{\surf}^{2}+2 q_{L}q_{N}-q_{N}^{2}\right)\right) + \mu_{\alpha}\left(k_{N}^{2}-k_{\surf}^{2}\right)  \left(k_{\surf}^{2}+q_{L} q_{N}\right)\right)}{c_{L,\alpha}^{2}}
\end{eqnarray}

For the reflected $L,N$-mixed waves, we have

\begin{eqnarray}
-\frac{c_{L,\alpha}}{2 i c_{N,\alpha} k_{N} k_{\surf}}\Delta R_{LN} & = & \left(k_{\surf}^{2}-k_{N}^{2}\right) \mu_{\alpha} \left(\frac{\omega^{2} \lambda_{\beta}}{c_{L,\beta}^{2}}+2 \mu_{\alpha} \left(k_{\surf}^{2}+q_{L} q_{N}\right)\right)\nonumber\\
& &+\mu_{\beta} \left(\frac{\omega^{2}\left(q_{N}^{2}-k_{\surf}^{2}\right) \lambda_{\beta}}{c_{L,\beta}^{2}}-2 \mu_{\beta} \left(q_{L}^{2} \left(k_{\surf}^{2}-q_{N}^{2}\right)-2 k_{\surf}^{2} q_{L} q_{N}\right)\right)\nonumber\\
& &-2 \mu_{\alpha}\mu_{\beta} \left(k_{N}^{2} q_{L} (q_{L}-q_{N})-k_{\surf}^{2} \left(-k_{\surf}^{2}+q_{L}^{2}-3 q_{L} q_{N}+q_{N}^{2}\right)\right),
\end{eqnarray}

\begin{eqnarray}
\frac{c_{N,\alpha}}{2 i c_{L,\alpha} k_{L} k_{\surf}}\Delta R_{NL} & = & \frac{\omega^{2} \lambda_{\alpha} \left(2 \mu_{\alpha} \left(k_{\surf}^{2}+q_{L} q_{N}\right)+\mu_{\beta} \left(-k_{\surf}^{2}-2 q_{L}q_{N}+q_{N}^{2}\right)\right)}{c_{L,\alpha}^{2}}-\frac{\omega^{2} \lambda_{\beta} \left(2 k_{\surf}^{2} \mu_{\alpha}+\left(q_{N}^{2}-k_{\surf}^{2}\right) \mu_{\beta}\right)}{c_{L,\beta}^{2}}\nonumber\\
& & + 2\mu_{\alpha} \mu_{\beta} \left(2 q_{L} q_{N} \left(k_{\surf}^{2}-k_{L}^{2}\right)+k_{L}^{2} \left(q_{N}^{2}-k_{\surf}^{2}\right)-2 k_{\surf}^{2} q_{L}^{2}\right)\nonumber\\
& & + 4 k_{L}^{2} \mu_{\alpha}^{2} \left(k_{\surf}^{2}+q_{L} q_{N}\right)+ 2 q_{L} \mu_{\beta}^{2} \left(q_{L} \left(k_{\surf}^{2}-q_{N}^{2}\right)-2 k_{\surf}^{2} q_{N}\right)
\end{eqnarray}

For the reflected $N$-polarized phonons, we find

\begin{eqnarray}
\Delta R_{NN} & = & -2 \Bigg[\mu_{\alpha} \mu_{\beta} \Big[\omega^{4} \left(\frac{k_{L} q_{N}}{c_{L,\beta}^{2} c_{N,\alpha}^{2}}-\frac{k_{N} q_{L}}{c_{L,\alpha}^{2} c_{N,\beta}^{2}}\right)\nonumber\\
& &+k_{\surf}^{2}\left(k_{L}^{2} \left(\left(k_{\surf}^{2}+2 q_{L} q_{N}-q_{N}^{2}\right)\right)+k_{L} k_{N} \left(k_{\surf}^{2}-2 q_{L}^{2}+4 q_{L}q_{N}-q_{N}^{2}\right)+q_{L} \left(k_{\surf}^{2}-k_{N}^{2}\right) (q_{L}-q_{N})\right)\Big]\nonumber\\
& &+k_{L} \mu_{\alpha}^{2} \left(q_{L} q_{N} + k_{\surf}^{2}\right)\left(k_{L}\left(k_{N}^{2}-k_{\surf}^{2}\right)-2 k_{N} k_{\surf}^{2}\right) +q_{L} \mu_{\beta}^{2} \left(k_{L} k_{N}-k_{\surf}^{2}\right)\left(q_{L} \left(k_{\surf}^{2}-q_{N}^{2}\right)-2 k_{\surf}^{2} q_{N}\right)\Bigg]\nonumber\\
& &-\frac{\omega^{2} \lambda_{\beta} \left(\mu_{\alpha} \left(k_{L} \left(\frac{q_{N} \omega^{2}}{c_{N,\alpha}^{2}}-2 k_{N} k_{\surf}^{2}\right)+k_{\surf}^{2} \left(k_{\surf}^{2}-k_{N}^{2}\right)\right)+\left(k_{\surf}^{2}-q_{N}^{2}\right) \mu_{\beta} \left(k_{L}k_{N}-k_{\surf}^{2}\right)\right)}{c_{L,\beta}^{2}}\nonumber\\
& &+\frac{\omega^{2} \lambda_{\alpha} \left(\mu_{\beta} \left(\frac{k_{N} q_{L} \omega^{2}}{c_{N,\beta}^{2}}+k_{\surf}^{2} \left( q_{N}^{2}-k_{\surf}^{2}-2q_{L} q_{N}\right)\right)+\left(k_{\surf}^{2}-k_{N}^{2}\right) \mu_{\alpha} \left(q_{L} q_{N} + k_{\surf}^{2}\right)\right)}{c_{L,\alpha}^{2}}
\end{eqnarray}

The various transmission matrix coefficients are also found,

\begin{eqnarray}
\frac{c_{L,\beta}}{2 c_{L,\alpha} k_{L}}\Delta T_{LL} & = & 2 \omega^{2} \mu_{\alpha} \left(\frac{k_{N} \left(2 k_{\surf}^{2} \mu_{\alpha}+\left(q_{N}^{2}-k_{\surf}^{2}\right) \mu_{\beta}\right)}{c_{L,\alpha}^{2}}+\frac{q_{N} \left(k_{L}^{2} \mu_{\alpha}+k_{\surf}^{2}   \mu_{\beta}\right)}{c_{N,\alpha}^{2}}\right)\nonumber\\
& &+\frac{\omega^{2} \lambda_{\alpha} \left(\mu_{\alpha} \left(\frac{q_{N} \omega^{2}}{c_{N,\alpha}^{2}}+2 k_{N} k_{\surf}^{2}\right)+k_{N}\left(q_{N}^{2}-k_{\surf}^{2}\right) \mu_{\beta}\right)}{c_{L,\alpha}^{2}}
\end{eqnarray}

\begin{eqnarray}
\frac{c_{L,\beta}}{2 i c_{N,\alpha} k_{N} k_{\surf}}\Delta T_{LN} & = & \omega^{2} \left(\frac{\left(\lambda_{\alpha}+2 \mu_{\alpha}\right) \left(\left(k_{N}^{2}-k_{\surf}^{2}\right) \mu_{\alpha}+\left(k_{\surf}^{2}-q_{N}^{2}\right) \mu_{\beta}\right)}{c_{L,\alpha}^{2}}+\frac{2k_{L} q_{N} \mu_{\alpha} \left(\mu_{\beta}-\mu_{\alpha}\right)}{c_{N,\alpha}^{2}}\right)
\end{eqnarray}

\begin{eqnarray}
\frac{c_{N,\beta}}{2 i c_{L,\alpha} k_{L} k_{\surf}}\Delta T_{NL} & = & 2 \omega^{2} \left(\frac{k_{N} q_{L} \left(\lambda_{\alpha}+2 \mu_{\alpha}\right) \left(\mu_{\beta}-\mu_{\alpha}\right)}{c_{L,\alpha}^{2}}+\frac{\mu_{\alpha} \left(k_{L}^{2} \mu_{\alpha}-q_{L}^{2} \mu_{\beta}\right)}{c_{N,\alpha}^{2}}\right)+\omega^{4} \mu_{\alpha} \left(\frac{\lambda_{\alpha}}{c_{L,\alpha}^{2} c_{N,\alpha}^{2}}-\frac{\lambda_{\beta}}{c_{L,\beta}^{2} c_{N,\alpha}^{2}}\right)
\end{eqnarray}

\begin{eqnarray}
\frac{c_{N,\beta}}{2 c_{N,\alpha} k_{N}}\Delta T_{NN} & = & \frac{\omega^{2}}{c_{N,\alpha}^{2}}\left(\frac{c_{N,\alpha}^{2} q_{L} \left(\lambda_{\alpha}+2 \mu_{\alpha}\right) \left(\left(k_{N}^{2}-k_{\surf}^{2}\right) \mu_{\alpha}+2 k_{\surf}^{2} \mu_{\beta}\right)}{c_{L,\alpha}^{2}}+\frac{k_{L} \omega^{2} \mu_{\alpha} \lambda_{\beta}}{c_{L,\beta}^{2}}+2 k_{L} \mu_{\alpha} \left(k_{\surf}^{2} \mu_{\alpha}+q_{L}^{2} \mu_{\beta}\right)\right)
\end{eqnarray}

The results we have shown above for the reflection and transmission of acoustic waves between two planar isotropic elastic media are consistent with the classical Knott's equations used in geology \cite{knott1899elastic}. 

%\end{widetext}

\subsection{Surface state for $\omega=0$}
Here we investigate in detail the $\omega=0$ limit of the reflection and transmission matrices. We find that the $M$ mode converges to a finite result, while the $R_{LN}$ matrix element, diverge as $\omega^{-2}$. In particular, we find
\begin{eqnarray}
R_{MM} = \frac{\mu_{\alpha}-\mu_{\beta}}{\mu_{\alpha}+\mu_{\beta}} + \mathcal{O}\left[\omega^{2}\right]
\end{eqnarray}
\begin{eqnarray}
R_{LN}(\omega) = \left(\frac{c_{L,\alpha} k_{\surf}}{\omega}\right)^{2}
\frac{4 \mu_{\alpha} \left(\mu_{\alpha}-\mu_{\beta}\right)}{\lambda_{\alpha} \left(\mu_{\alpha}+\mu_{\beta}\right)+\mu_{\alpha} \left(\mu_{\alpha}+3 \mu_{\beta}\right)}\left(
\begin{array}{cc}
 1 & -\frac{c_{N,\alpha}}{c_{L,\alpha}} \\
 -\frac{c_{L,\alpha}}{c_{N,\alpha}} & 1 \\
\end{array}
\right) + \mathcal{O}\left[\omega^{0}\right]
\end{eqnarray}

Usually, a divergence in the reflection matrix signals the appearance of a bound state, that we would expect to be double-degenerate at $\omega=0$. To find such a bound state, we start from the scattering problem ansatz given in \Eq{Scattering_Ansatz} applied to the boundaries conditions given \Eq{No_Slip_BC_Appendix} and \Eq{HidrostaticEqBC_expanded} by taking $A_{i,P}=0$ since by definition there is no incident wave in this case. The reflected and transmitted waves at $\omega=0$ are exponential decaying functions: the reflected wave decays as $e^{k_{\surf}z} = e^{-k_{\surf}\abs{z}}$ in $\alpha$ medium ($z<0$) and the transmitted wave decays as $e^{-k_{\surf}z} = e^{-\abs{k_{\surf}}z}$ in $\beta$ medium ($z>0$).

At $\omega=0$, the scattering problem for $L$ and $N$ modes transforms into two copies of the same equality (indicating a double degeneration)
\begin{eqnarray}
2k_{\surf}^{2}\left(c_{L,\alpha}A_{L,r} + c_{N,\alpha}A_{N,r}\right)
\left(\begin{array}{c}
1\\
1\\
2\mu_{\alpha}k_{\surf}\\
2\mu_{\alpha}k_{\surf}
\end{array}\right) + \mathcal{O}\left[\omega^{2}\right] = \left(\begin{array}{c}
0\\
0\\
0\\
0
\end{array}\right) + \mathcal{O}\left[\omega^{2}\right],
\end{eqnarray}
which yields the following relationship
\begin{eqnarray}
A_{N,r} = - \frac{c_{L,\alpha}}{c_{N,\alpha}}A_{L,r}.
\end{eqnarray}
This means that the bound state only exists in $z<0$ region corresponding to the $\alpha$ material.

The bound state can then be written as
\begin{eqnarray}
\ket{\psi_{\alpha}(\bm{r})} = \lim_{\omega\to 0}A_{N,r}\left[ -\frac{c_{N,\alpha}}{c_{L,\alpha}}\bm{L}_{\alpha}(\bm{k}_{L,r},\bm{r}) + \bm{N}_{\alpha}(\bm{k}_{N,r},\bm{r})\right]\Theta(-z),
\end{eqnarray}
with $\dlim_{\omega\to0}\bm{k}_{L,r} = \lim_{\omega\to0}\bm{k}_{N,r} = (k_{x},k_{y},\ii k_{\surf})$. The orthonormalization of this state imposes
\begin{eqnarray}
A_{N,r} = -\frac{2 \sqrt{2} c_{L,\alpha}^{2}c_{N,\alpha}k_{\surf}^{3/2}}{\omega  \sqrt{c_{L,\alpha}^{4} + c_{N,\alpha}^{4}}},
\end{eqnarray}
which finally yields
\begin{eqnarray}
\ket{\psi_{\alpha}(\bm{r})} = e^{\ii(k_{x} x+k_{y} y)}e^{k_{\surf} z}\sqrt{\frac{2k_{\surf}}{c_{L,\alpha}^{4}+c_{N,\alpha}^{4}}}
\left[
\left(\begin{array}{c}
\ii\cos(\varphi)c_{L,\alpha}^{2}\\
\ii\sin(\varphi)c_{L,\alpha}^{2}\\
-c_{N,\alpha}^{2}
\end{array}\right) + k_{\surf} z \left(c_{L,\alpha}^{2}-c_{N,\alpha}^{2}\right)
\left(\begin{array}{c}
\ii\cos (\varphi )\\
\ii\sin (\varphi )\\
1
\end{array}\right)
\right]\Theta(-z)
\end{eqnarray}

In a similar way, we find the bound state that can only exist in the $z>0$ region corresponding to the $\beta$ material. By using  a factorization basis with $k_{z}$ \textit{in the opposite direction compared to the previous case}, the corresponding normalized wave function is obtained as 
\begin{eqnarray}
\ket{\psi_{\beta}(\bm{r})} = e^{\ii(k_{x} x+k_{y} y)}e^{-k_{\surf} z}\sqrt{\frac{2k_{\surf}}{c_{L,\beta}^{4}+c_{N,\beta}^{4}}}
\left[
\left(\begin{array}{c}
\ii\cos(\varphi)c_{L,\beta}^{2}\\
\ii\sin(\varphi)c_{L,\beta}^{2}\\
c_{N,\beta}^{2}
\end{array}\right) + k_{\surf} z \left(c_{N,\beta}^{2}-c_{L,\beta}^{2}\right)
\left(\begin{array}{c}
\ii\cos (\varphi )\\
\ii\sin (\varphi )\\
-1
\end{array}\right)
\right]\Theta(+z)
\end{eqnarray}
By construction, the so-obtained bound states are orthonormal 
 $\braket{\psi_{\alpha}(\bm{r})}{\psi_{\beta}(\bm{r})} = \delta_{\nu\mu}$ and they are eigenstates of the elastic Hamiltonian 
 $\hat{H}\ket{\psi_{\mu}(\bm{r})} = \dlim_{\omega\to0}\hbar\omega\ket{\psi_{\mu}(\bm{r})} = 0$ and they are double degenerate. 
Also by construction, these bound states fulfil the no-slip boundary condition, but the hidrostatic boundary condition in \Eq{HidrostaticEqBC_expanded},  is fulfilled only by the following orthonormal bound state 
\begin{eqnarray}
\ket{B(z)} = \frac{C_{\alpha}\ket{\psi_{\alpha}(z)} - C_{\beta}\ket{\psi_{\beta}(z)}}{C_{\alpha}^{2} + C_{\beta}^{2}},
\end{eqnarray}
with
\begin{eqnarray}
C_{\alpha} = \mu_{\alpha}\frac{\sqrt{ c_{L,\alpha}^{4} + c_{N,\alpha}^{4} }}{ c_{L,\alpha}^{2} - c_{N,\alpha}^{2} },
\hspace{2cm}
C_{\beta} = \mu_{\beta}\frac{\sqrt{ c_{L,\beta}^{4} + c_{N,\beta}^{4} }}{ c_{L,\beta}^{2} - c_{N,\beta}^{2} }
\end{eqnarray}
We note that this (double degenerated) state exists regardless of the presence of other objects with respect to the boundary between objects $\alpha$ and $\beta$. Thus its possible contribution to the Casimir energy or force will be independent of the relative distances between the considered objects and, as a consequence, it makes no logical sense to consider its contribution as part of the Casimir interaction between interacting objects. This argument justifies that the $\omega=0$ contribution to the Casimir energy of the Lifshitz formula for the $L$ and $N$ modes should not be considered, and the appearance of a double pole in the phononic reflection matrix coming from the double degeneration of the bound state.

\subsection{Basic Materials Properties and Representative Examples of the Phononic Casimir Interaction}

In Table \ref{Fig_Props_Materials_complete}, we gather data for some of the basic properties used in the calculations for Fig. 2 in the main text. 

\begin{table}[H]
\caption{ Basic properties of different materials as found in their respective references: mass density $\rho$ in $kg/m^{3}$, the Lamé parameters $\lambda=C_{12}$, $\mu=C_{44}$ and $C_{11}$ given in $\text{ GPa} = 10^{9}kg/(m s^{2})$, the sound velocities $c_{\ell}=\sqrt{\dfrac{C_{11}}{\rho}}$ and $c_{t}=\sqrt{\dfrac{\mu}{\rho}}$ expressed in $m/s$ and static dielectric constant $\varepsilon (0)$ For hexagonal and wurzite BN, the average dielectric constant is estimated as $\varepsilon (0) = \sqrt{\varepsilon_{\parallel} \varepsilon_{\perp}}$. Since indium (In) is a metal, its dielectric constant is considered as infinity.}
\begin{center}
\begin{ruledtabular}
\begin{tabular}{|c|c|c|c|c|c|c|c|c|}
\hline
Material & $\rho\,(\text{kg}/\text{m}^{3})$ & $\lambda\,(\text{GPa})$ & $\mu\,(\text{GPa})$ & $C_{11}\,(\text{GPa})$ & $c_{\ell}\,(\text{m}/\text{s})$ & $c_{t}\,(\text{m}/\text{s})$ & $\varepsilon (0)$ & Reference\\ \hline
Ge      & 5323 & 44    & 66.7  &  126 & 4865.27 & 3539.85 & 16.2 & \cite{Levinshtein1996handbook}\\ \hline
Si      & 2329 & 64    & 79.6  &  166 & 8442.47 & 5846.17 & 11.7 & \cite{Levinshtein1996handbook}\\ \hline
Diamond & 3514 & 124   & 578   & 1070 & 17447.3 & 12823 & 5.7 & \cite{Levinshtein1996handbook}\\ \hline
BN$_{\text{cub}} ($\text{cubic}$)$      & 3487 & 190   & 480   &  820 & 15334.9 & 11732.6 & 7.1 & \cite{Levinshtein2001handbook}\\ \hline
BN$_{\text{hex}} ($\text{hexagonal}$)$     & 2180	&30 &	15	& 75 & 18548.2 & 8295 & 5.9 & \cite{Levinshtein2001handbook}\\ \hline
BN$_{\text{w}} ($\text{wurtzite}$)$      & 3487&	13.4&	38.8&	98.2&	16781.5&	10548.5 & 5.9 & \cite{Levinshtein2001handbook}\\ \hline
In      & 7300 & 39.50 & 6.55 & 44.50 & 2467.30 & 946.59 & $ \infty$ & \cite{Haynes2016handbook} \\ \hline
% &  &  &  &  &  &  & \\
\end{tabular}
\end{ruledtabular}
\label{Fig_Props_Materials_complete}
\end{center}
\end{table}

In \Fig{Supporting_Fig_2}, results from the numerical calculations of the different contributions of phonon polarizations to the Casimir effects are shown at room temperature for three representative cases: Ge/Si/Ge, Diamond/Si/Diamond, and Diamond/BN$_{hex}$/Diamond. The standard Casimir energy from electromagnetic fluctuations in the quantum mechanical and thermal limits are also shown for comparison. These are calculated using the Lifshitz approach \cite{lifshitz1956molecular, RevModPhys.88.045003},
\begin{eqnarray}
E_{em}^{qm} &\approx & - \frac{\hbar c}{16 \pi^2 d^3} \int_0^1 dx \left\{ \text{Li}_4 \left( \left[ \frac{\varepsilon_0 - \varepsilon_1 \sqrt{1 + \left(\frac{\varepsilon_1}{\varepsilon_0} - 1 \right) x^2} }{\varepsilon_0 + \varepsilon_1 \sqrt{1 + \left(\frac{\varepsilon_1}{\varepsilon_0} - 1 \right) x^2} } \right]^2 \right) + \text{Li}_4 \left( \left[ \frac{1- \sqrt{1 + \left(\frac{\varepsilon_1}{\varepsilon_0} - 1 \right) x^2}}{1 + \sqrt{1 + \left(\frac{\varepsilon_1}{\varepsilon_0} - 1 \right) x^2} } \right]^2 \right)\right\},\\
E_{em}^{T} &\approx & - \frac{k_B T}{16 \pi d^2} \text{Li}_3 \left( \left[ \frac{\varepsilon_0 - \varepsilon_1}{\varepsilon_0 + \varepsilon_1} \right]^2 \right),
\end{eqnarray}
where $\varepsilon_0$ and $\varepsilon_1$ are static dielectric constant at zero frequency of Materials $0$ and $1$, as shown in the schematics in Fig. 1 of the main text. 

We find that the dominant contribution from phonons at separations larger than an Angstrom comes from the thermal energy associated with the $M$-polarization. The electromagnetic Casimir interaction, however, is primarily of quantum mechanical origin for the shown distance range. For many cases, the standard Casimir effect dominates over the Phononic analog, however, for some materials it is possible the $\mathcal{E}_M^T$ overtakes  $E_{em}^{qm}$, as is the case for Diamond plates separated by a gap filled with hexagonal BN.

\begin{figure}[H]
\centering
\includegraphics[width=0.95\linewidth]
{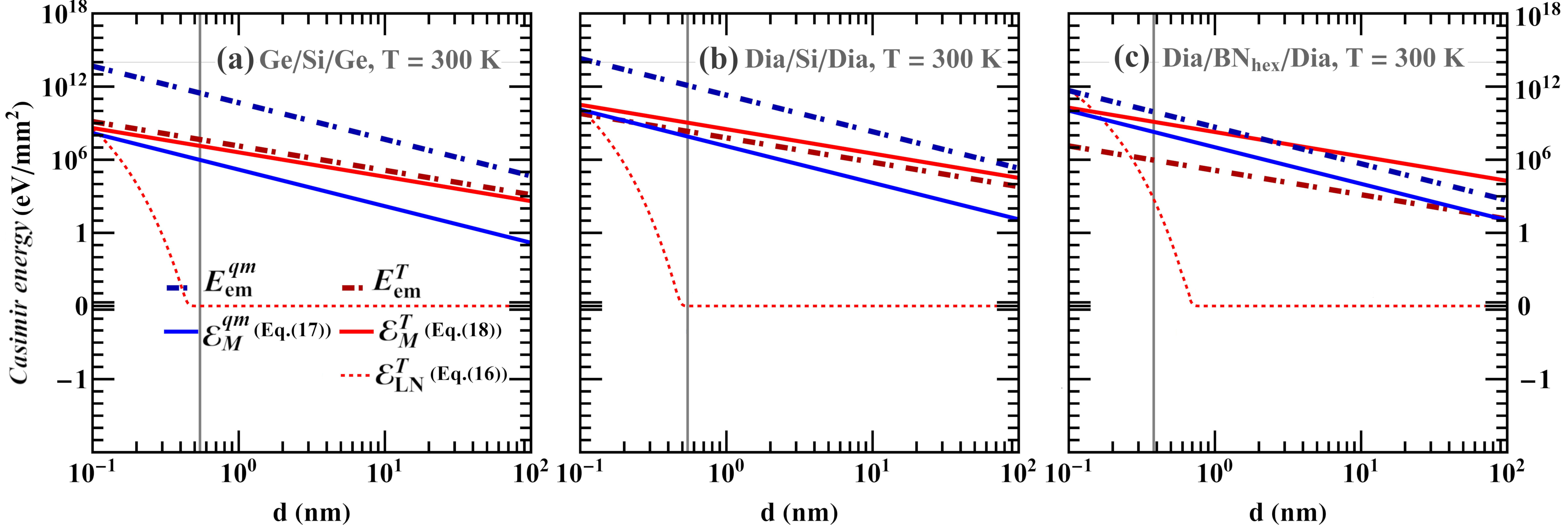}
\caption{Phononic and electromagnetic Casimir interaction energies for (a) Ge/Si/Ge, (b) Diamond/Si/Diamond, and (c) Diamond/BN$_{\text{hex}}$/Diamond configurations. The contributions from the quantum mechanical $\mathcal{E}_M^{qm}$ and the thermal $\mathcal{E}_M^{T}$ and $\mathcal{E}_{L,N}^{T}$ limits in Eqs. (17), (18), and (16) from the main text, respectively, are shown. Similarly, contributions from the quantum mechanical $E_{em}^{qm}$ and thermal $E_{em}^T$ limits are also shown.}
\label{Supporting_Fig_2}
\end{figure}

Furthermore, we can estimate the ratio between the Phononic and electromagnetic Casimir interactions in the range of $d \leq 100 \text{ nm}$ and $T \leq 300 \text{ K}$. Indeed, in this case, the quantum limit dominates the electromagnetic Casimir energy i.e. $E_{em}^{qm} \gg E_{em}^T$ \cite{RevModPhys.88.045003}. For dielectrics, the quantum limit is typically determined by p-type electromagnetic waves further expressed as
\begin{eqnarray}
E_{em} \approx E_{em}^{qm} && \approx - \frac{\hbar c}{16 \pi^2 d^3} \int_0^1 \text{Li}_4 \left( \left[ \frac{\varepsilon_0 - \varepsilon_1 \sqrt{1 + (\frac{\epsilon_1}{\epsilon_0}-1) x^2} }{\varepsilon_0 + \varepsilon_1 \sqrt{1 + (\frac{\epsilon_1}{\epsilon_0}-1) x^2} } \right]^2 \right) dx \approx - \frac{\hbar c}{16 \pi^2 d^3} \text{Li}_4 \left( \left[ \frac{\varepsilon_0 - \varepsilon_1 }{\varepsilon_0 + \varepsilon_1} \right]^2 \right) \nonumber\\
                                 && \approx - \frac{\hbar c}{16 \pi^2 d^3}  \left( \frac{\varepsilon_0 - \varepsilon_1 }{\varepsilon_0 + \varepsilon_1} \right)^2 .
 \end{eqnarray}
The last line holds if the difference of static dielectric constant of two materials is small enough $\left| \varepsilon_0 - \varepsilon_1 \right| \ll  \left(\varepsilon_0 + \varepsilon_1\right)$.

Phononic Casimir interaction, on the other hand, is dominated by the high temperature limit of the M mode polarization, therefore
\begin{eqnarray}
\mathcal{E}_{ph} \approx  \mathcal{E}_{M}^{T} && = - \frac{k_B T}{16 \pi d^2} \text{Li }_3 \left( \left[ \frac{\mu_0 - \mu_1 }{\mu_0 + \mu_1} \right]^2 \right) \approx - \frac{k_B T}{16 \pi d^2} \left( \frac{\mu_0 - \mu_1 }{\mu_0 + \mu_1} \right)^2 .
\end{eqnarray}
The last line holds if the difference of Lam\'e parameters of two materials is small enough $\left| \mu_0 - \mu_1 \right| \ll  \left(\mu_0 + \mu_1\right)$.

Combining these approximations, the ratio between phononic and electromagnetic Casimir energy per area is given as
\begin{eqnarray}
\dfrac{\mathcal{E}_{ph.}}{E_{em}} \approx  \frac{k_B T d}{\pi \hbar c} \left( \frac{(\mu_0 - \mu_1)/(\varepsilon_0 - \varepsilon_1) }{(\mu_0 + \mu_1)/(\varepsilon_0 + \varepsilon_1)} \right)^2,
\end{eqnarray}
which well agree with numerical results showed in Fig. 2.

Finally, Fig. \ref{Supporting_Fig_3} shows the  magnitude of the unitless constant $B$ for the Phononic Casimir energy from the $L, N$ polarization modes in Eqs. (15, 16) in the main text. The results from the calculations indicate that it varies in the $(0.01, 10)$ range for the materials combinations considered here.

\begin{figure}[H]
\centering
\includegraphics[width=0.6\linewidth]
{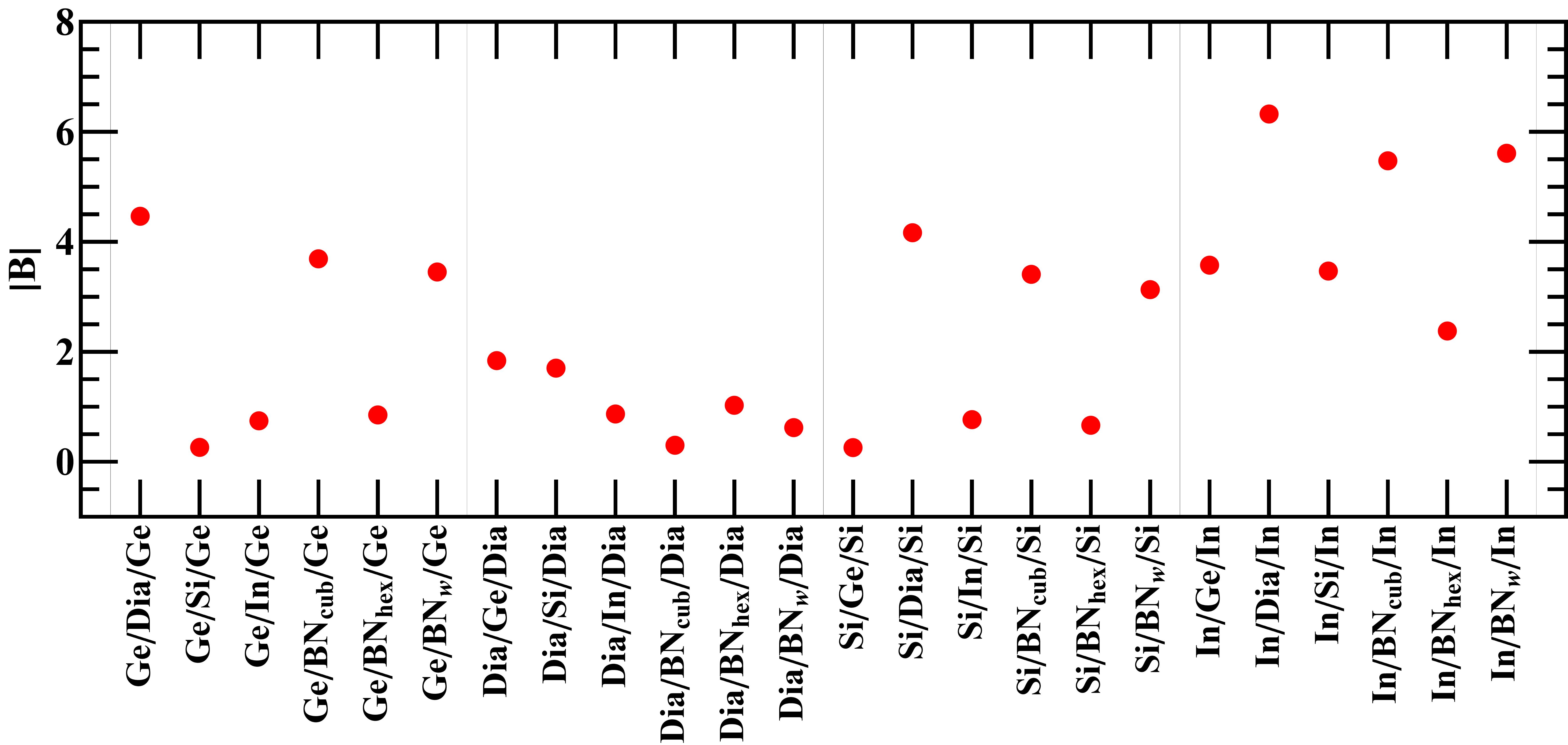}
\caption{Typical values of the magnitude of the unitless constant $B$ in Eqs (15, 16) in the main text.}
\label{Supporting_Fig_3}
\end{figure}

\subsection{Proposed experimental observation of the phononic Casimir interaction}

We propose that the Phononic Casimir interaction may be observed by measuring the Young's modulus of thin films by using mechanically and thermally controlled transducer set-up \cite{Kellogg2005,MORENO2021}. By placing Material 0 between two substrates made of Material 1 (Figure 1 in manuscript), one can generate stress-strain data at different temperatures to obtain the distance-dependent Young’s modulus $\Xi(d)$, which depends on the pressure and properties of the Material 1/Material 0 interfaces. 

For bulk elastic Material 0, the stress $\sigma$ depends on the strain $\epsilon$ and the measured $\sigma(\epsilon)$ is used to obtain the bulk Young's modulus $\Xi^{bulk}$. For this purpose, for example, one applies external tension $T_x^{appl}$ along the $x$-axis  $T_x^{appl}$ in the quasi-static regime $\rho\frac{D v_{x}}{D t}\approx 0$ (see Eq. 1 in the main text). In this way, one obtains $\varepsilon_{xx} = \partial_{x}u_{x}$ in the material, such that 
\begin{eqnarray}
T_{x}^{\text{applied}} = T_{x}^{\text{total}} = \sigma_{xx}(\varepsilon) = \Xi^{bulk}\varepsilon_{xx}.
\end{eqnarray}

In the case of a thin elastic material of thickness $d$, the total stress $T_{x}^{\text{total}}$ must be modified by the interface Casimir pressure $p_{C} = - \frac{1}{A}\frac{\partial E_{C}}{\partial d} = - \frac{k_{B}T}{8\pi d^{3}}\text{Li}_{3}\left(\left[\frac{\mu_{0} - \mu_{1}}{\mu_{0} + \mu_{1}}\right]^{2}\right)$ ($A$ - unit area).

-------------------------------------

%As a consequence of the presence of the substrates of Material 1, a phononic Casimir pressure that depends on the elastic properties of the interacting materials will appear. Note that, if we do not have the substrates of Material 1, the vacuum also behaves as a material with $\mu = \lambda = 0$ and another different phononic Casimir force will appear over the boundaries of Material 0.

Therefore, the phononic Casimir pressure will be an additional source of stress to the Material 0, proportional to the temperature, and inversely proportional to the thickness, because the additional Casimir pressure will be very well approached by
\begin{eqnarray}
p_{C} = - \frac{1}{A}\frac{\partial E_{C}}{\partial d} = - \frac{k_{B}T}{8\pi d^{3}}\text{Li}_{3}\left(\left[\frac{\mu_{0} - \mu_{1}}{\mu_{0} + \mu_{1}}\right]^{2}\right),    
\end{eqnarray}
%I expect that modifying $\mu_{\alpha}$ from $0$ (the case of vacuum) to a case where $\mu_{\alpha} \neq 0$ (with a thick material on top of the ''medium''), will modify the phononic Casimir pressure $p_{C}$.

If for a thick volume of the elastic material $0$ we have that the stress depends on the strain as $\sigma = \sigma(\varepsilon)$, in a stress-strain experiment, when we apply an external tension along the x-axis $T_{x}^{\text{applied}} = \frac{\partial F_{x}}{\partial A}$ at the quasi-static regime $\rho\frac{D v_{x}}{D t}\approx 0$, we obtain a strain $\varepsilon_{xx} = \partial_{x}u_{x}$ in the material in such a way that 
\begin{eqnarray}
T_{x}^{\text{applied}} = T_{x}^{\text{total}} = \sigma_{xx}(\varepsilon) = E\varepsilon_{xx},
\end{eqnarray}
where $E$ is the Young modulus. On the other side, when we have a thin elastic material of thickness $d$, even at rest there is a phononic Casimir pressure, therefore, the full mechanical stress over the material $T_{x}^{\text{total}}$ will be the external tension $T_{x}^{\text{applied}}$ in addition to the Casimir pressure $p_{C}(d) = p_{C}(d(1+\epsilon_{xx}))$, and we have
\begin{eqnarray}
T_{x}^{\text{total}} = T_{x}^{\text{applied}} - p_{C}(d(1+\epsilon_{xx})) = \sigma_{xx}(\varepsilon) = E\varepsilon_{xx},
\end{eqnarray}
The minus in $- p_{C}(d)$ comes from the fact that a compressive pressure should compress the system, therefore, we should have $\varepsilon_{xx}<0$ when $T_{x}^{\text{applied}}=0$. From this formula, in absence of external applied tension ($T_{x}^{\text{applied}}=0$), the thin film has \textit{less} thickness as would be expected if there would not be Casimir pressure. In particular, when $T_{x}^{\text{applied}}=0$, we have
\begin{eqnarray}
T_{x}^{\text{total}} = - p_{C}(d_{0}(1+\epsilon_{xx}^{(0)})) = \sigma_{xx}(\varepsilon^{(0)}) = E\varepsilon_{xx}^{(0)},
\end{eqnarray}
therefore, in absence of external applied stress, we have
\begin{eqnarray}
\sigma_{xx}(\varepsilon^{(0)}) + p_{C}(d_{0}(1+\epsilon_{xx}^{(0)})) = 0,
\end{eqnarray}
where $\epsilon_{xx}^{(0)}$ is the (small) internal strain that compress our plate from the expected initial thickness $d_{0}$ to the actual thickness $d_{1}=d_{0}(1+\epsilon_{xx}^{(0)})$. In addition to that, in stress-strain experiments, we apply an external stress or tension $T_{x}^{\text{applied}}$ and measure a strain $\epsilon_{xx}$ to a thin plate of thickness $d_{1} = d_{0}(1 + \epsilon_{xx}^{(0)})$ at rest, therefore, in our set-up, reordering terms we have
\begin{eqnarray}
T_{x}^{\text{applied}} = E(\epsilon_{xx}^{(0)} + \epsilon_{xx}) + p_{C}(d_{1}(1 + \epsilon_{xx})).
\end{eqnarray}
If we remain in the linear regime, we can expand in $\epsilon_{xx}$ this expression to obtain
\begin{eqnarray}
T_{x}^{\text{applied}} = E(\epsilon_{xx}^{(0)} + \epsilon_{xx}) + p_{C}(d_{1}) + d_{1}p'_{C}(d_{1}) \epsilon_{xx} + \mathcal{O}\left[\epsilon_{xx}^{2}\right].
\end{eqnarray}
Having into account that $d_{1} = d_{0}(1 + \epsilon_{xx}^{(0)})$, as we have seen before
\begin{eqnarray}
E\epsilon_{xx}^{(0)} + p_{C}(d_{1}) = 0,
\end{eqnarray}
and we obtain that the measured strain $\epsilon_{xx}$ when an external tension $T_{x}^{\text{applied}}$ is applied is
\begin{eqnarray}
T_{x}^{\text{applied}} = \Big[E + d_{1}p_{C}'(d_{1})\Big]\epsilon_{xx} + \mathcal{O}\left[\epsilon_{xx}^{2}\right],
\end{eqnarray}
it is to say, the presence of the Casimir pressure \textit{modifies} the Young modulus measured in a stress-strain experiment for small thickness $d_{1} = d_{0}(1 + \epsilon_{xx}^{(0)})$ from $E$ to
\begin{eqnarray}
E^{\text{exp}}(d_{1}) = E + d_{1}p'_{C}(d_{1}) = E + 3\frac{k_{B}T}{8\pi d_{1}^{3}}\text{Li}_{3}\left(\left[\frac{\mu_{0} - \mu_{1}}{\mu_{0} + \mu_{1}}\right]^{2}\right) = E - 3p_{C}(d_{1}).
\end{eqnarray}
From this result, we see that the experimentally measured Young modulus of thin plates should be linearly proportional to the temperature, depends on the elastic properties of the involved materials ($\mu_{0}$ and $\mu_{1}$) and, the thinner the material is, the greater $E^{\text{exp}}(d_{1})$ becomes.

\end{widetext}

\end{document}